\documentclass[pdflatex,sn-nature]{sn-jnl}

\usepackage{graphicx}
\usepackage{amsmath,amssymb,amsfonts}
\usepackage{xcolor}
\usepackage{xspace}
\usepackage{booktabs}
\usepackage{comment}
\usepackage{multirow}

\newcommand{\ha}{H$\alpha$\xspace}
\newcommand{\hb}{H$\beta$\xspace}

\newcommand{\paa}{Pa$\alpha$\xspace}
\newcommand{\hi}{H\,\textsc{i}\xspace}
\newcommand{\NHItwo}{\ifmmode N_{\rm HI,\mathit{n}=2}\else $N_{\rm HI,\mathit{n}=2}$\xspace\fi}
\newcommand{\kms}{\ifmmode\,{\rm km\,s^{-1}}\else\,km\,s$^{-1}$\xspace\fi}
\newcommand{\cmtwo}{\ifmmode\,{\rm cm^{-2}}\else\,cm$^{-2}$\xspace\fi}

\unnumbered  

\begin{document}

\title[Multiphase Gas in LRDs]{Multiphase Gas Structure in JWST Little Red Dots}

\author*[1]{\fnm{Seok-Jun} \sur{Chang}}\email{sjchang@mpa-garching.mpg.de} \author[2]{\fnm{Max} \sur{Gronke}} \author[3]{\fnm{Jorryt} \sur{Matthee}} \author[4,5]{\fnm{Charlotte} \sur{Mason}} 

\affil*[1]{\orgname{Max-Planck-Institut f\"{u}r Astrophysik}, \orgaddress{\street{Karl-Schwarzschild-Stra\ss e 1}, \postcode{85748}, \city{Garching b. M\"{u}nchen}, \country{Germany}}} \affil[2]{\orgname{Astronomisches Rechen-Institut, Zentrum f\"{u}r Astronomie, Universit\"{a}t Heidelberg}, \orgaddress{\street{M\"{o}nchhofstra\ss e 12-14}, \postcode{69120}, \city{Heidelberg}, \country{Germany}}} \affil[3]{\orgname{Institute of Science and Technology Austria (ISTA)}, \orgaddress{\street{Am Campus 1}, \postcode{3400}, \city{Klosterneuburg}, \country{Austria}}} \affil[4]{\orgname{Cosmic Dawn Center (DAWN)}} \affil[5]{\orgname{Niels Bohr Institute, University of Copenhagen}, \orgaddress{\street{Jagtvej 128}, \postcode{DK-2200}, \city{Copenhagen}, \country{Denmark}}}

\abstract{JWST has uncovered a population of compact high-redshift galaxies, `little red dots' (LRDs), exhibiting broad Balmer emission lines that often show narrow absorption features and strong Balmer breaks --- spectral signatures attributed to dense gas surrounding rapidly growing black holes, but whose small-scale structure remains unknown. A striking feature of LRD spectra is that the H$\beta$ absorption is systematically deeper than H$\alpha$, yet in any homogeneous gas this ordering is physically impossible. Here we model the H$\alpha$ and H$\beta$ line profiles using radiative transfer simulations for 11 LRDs with JWST sensitive high resolution spectroscopy, and independently constrain the $n=2$ hydrogen column density, \NHItwo, from the Balmer-break continuum. We find that the \NHItwo from H$\alpha$ in a homogeneous medium cannot reproduce the observed H$\beta$ absorption in several objects, and that this \NHItwo typically falls two orders of magnitude below the lower bound set by the Balmer break.
Both tensions are simultaneously resolved by a multiphase medium of dense clumps within diffuse hot ionized gas. This multiphase gas structure is a necessary ingredient for any viable LRD model, regardless of the nature of the central engine. 
}


\maketitle

JWST has revealed compact sources at $z \sim 4$--9, `little red dots' (LRDs), with red optical continua, broad Balmer emission lines of several thousand \kms, Balmer breaks, and Balmer absorption \cite{Matthee2024,Greene24,Kocevski23,Kokorev2024LRD,Setton24,Juodzbalis2024a,Labbe24,Hviding2025}, the latter seen in $\sim$60\% of LRDs with deep spectra \cite{Matthee2026,Juodzbalis2026,Davis2026}. They are widely interpreted as rapidly accreting black holes in dense gas, but their nature remains debated: models invoke dense gas cocoons \cite{Rusakov2025,Inayoshi2025,Kido2025,Sneppen2026,Naidu2026}, super-Eddington accretion \cite{MadauMaiolino2026,Chon2026a}, and dust-free black-hole atmospheres \cite{Naidu2025,deGraaff2025}.
In most of these scenarios, scattering of Balmer lines through dense gas plays a central role in shaping the observed spectra \cite{ChangScattering2026}, but thus far, observed Balmer line profiles have not been confronted with radiative transfer models with scattering physics.

Interpreting these absorption features requires a model for how Balmer photons propagate through the gas. Previous LRD analyses fit the profiles with multi-component parametric functions \cite{Matthee2024,Rusakov2025,Torralba2026,dEugenio2026} or with Sobolev-approximation line transfer \cite{Sneppen2026}, both assuming simple absorption or moderate optical depth.
At the column densities implied by the Balmer break ($\NHItwo \gtrsim 10^{17}$\cmtwo), however, the line-center optical depth of the Balmer lines exceeds $10^3$: photons scatter thousands of times before escaping, and their profiles depend on the gas geometry, frequency redistribution, and branching between decay channels.
We previously modeled the full scattering behavior of the Balmer lines in LRDs with the Monte Carlo radiative-transfer code \textsc{rt-scat}\cite{ChangScattering2026}. 
Here we apply it to fit the observed H$\alpha$ and H$\beta$ profiles of LRDs.

Balmer absorption and Balmer-break strength are empirically correlated across the LRD population\cite{Matthee2026}, and physically linked: H$\alpha$, H$\beta$, and the Balmer continuum all arise from hydrogen in the $n=2$ level.                    
The $n=2$ column density, \NHItwo, therefore sets all three, so they form a single consistency test rather than independent diagnostics. This constraint has not been imposed in previous modeling of LRD spectra. 
We therefore start from the simplest assumption that the same $n=2$ \hi gas produces all three features, and ask what structure this gas must have.

\begin{figure*}[ht]
  \centering
  \includegraphics[width=\textwidth]{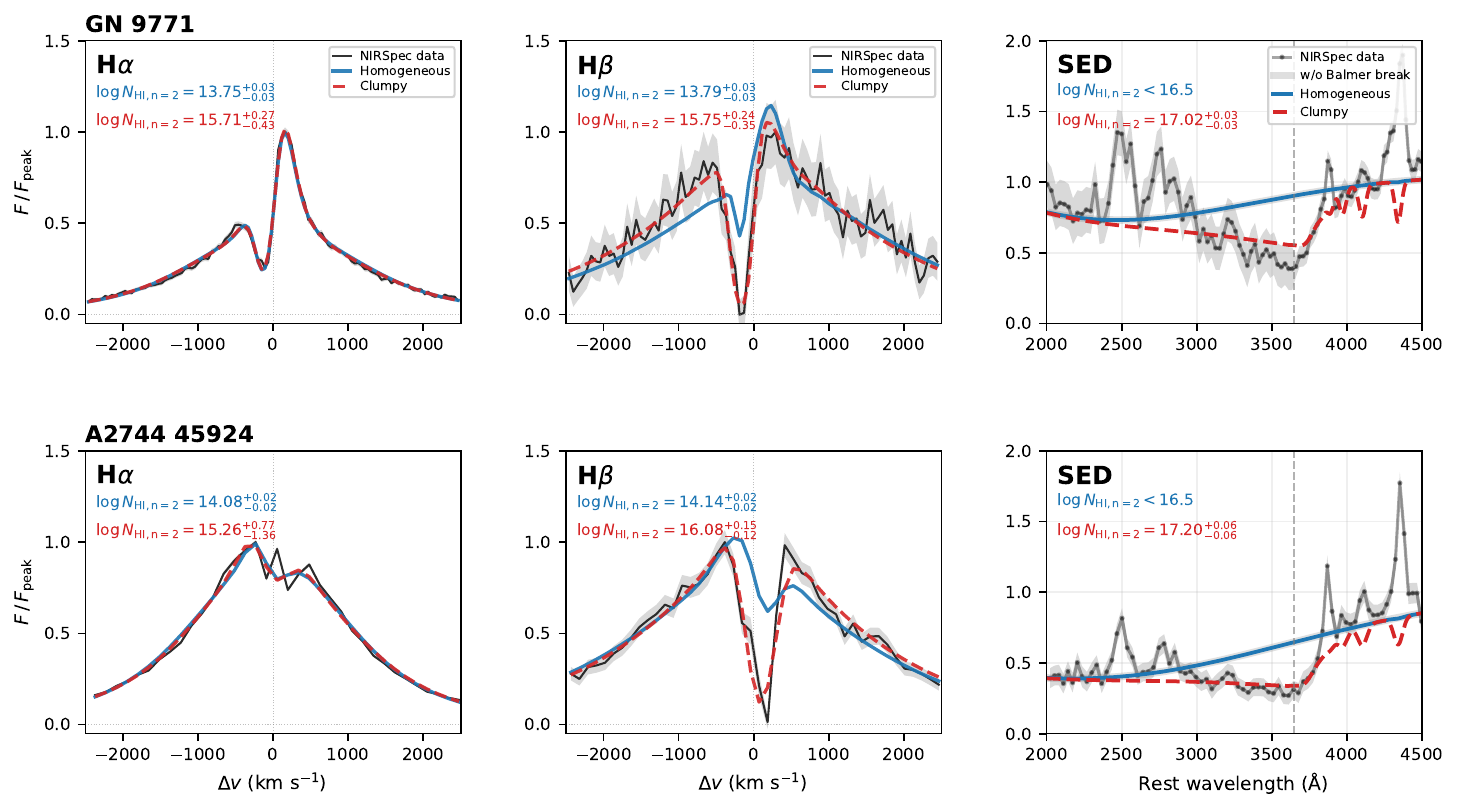}
  \caption{\textbf{H$\alpha$, H$\beta$, and Balmer-break fits for two representative
  LRDs.} The top and bottom panels show fitting results of GN~9771 and A2744~45924, respectively. The columns show H$\alpha$ (left), H$\beta$ (middle), and the rest-frame SED around the Balmer break (right). 
  Black curves represent the observational data, NIRSpec for H$\alpha$ \& H$\beta$ and PRISM for the SED; grey shading indicates the $\pm1\sigma$ uncertainty. 
  The blue solid and red dashed lines are the best fits for the homogeneous and clumpy models, respectively. Fit in two steps: H$\alpha$ alone first, then H$\beta$ and the Balmer-break SED fits with \NHItwo prior from H$\alpha$ fit (see
  Methods). The median \NHItwo with its 16th--84th percentile range is annotated in each panel. Homogeneous SED fits that cannot produce the Balmer break (84th percentile of \NHItwo below $10^{16.5}$\cmtwo, i.e. Balmer-continuum optical depth $\tau_{\rm BB} < 0.5$) are shown without uncertainties and labeled $\log N_{\rm HI,\mathit{n}=2} < 16.5$ (see Methods).
  For both objects, \NHItwo from H$\alpha$ in the homogeneous model are too low to
  reproduce the observed H$\beta$ or the Balmer break, while the clumpy model provides 
  \NHItwo values of H$\alpha$ consistent with both. Fits for the remaining 9 objects are shown in Extended Data Figures~\ref{fig:ext_fit1} and \ref{fig:ext_fit2}.}
  \label{fig:example_fit}
\end{figure*}

Applying this test reveals a puzzle. In LRD spectra, the H$\alpha$ absorption trough is typically \textit{weaker} than that of H$\beta$ \cite{dEugenio2026,Matthee2026}, which is impossible in a homogeneous gas: the H$\alpha$ line cross-section is $\approx 5.7$ times larger than that of H$\beta$\cite{ChangScattering2026}, so at the same \NHItwo, H$\alpha$ absorption is stronger than H$\beta$, or both lines are saturated.
Moreover, homogeneous Balmer-line fits give $\NHItwo \sim 10^{14}$\cmtwo, three orders of magnitude below the $\gtrsim 10^{17}$\cmtwo required by the Balmer break (left panels of Figure~\ref{fig:example_fit}). No single \NHItwo in a homogeneous medium satisfies all three observables.

We apply this test to the 11 LRDs from the broad-line sample of \cite{Matthee2026} with high signal-to-noise NIRSpec spectra of both H$\alpha$ and H$\beta$. A homogeneous single-phase gas fails it, whereas a multiphase medium of optically thick $n=2$ \hi clumps largely removes the line--break \NHItwo tension. This line--break comparison relies on the shared-gas assumption, but the multiphase model also reproduces H$\alpha$ and H$\beta$ at a consistent \NHItwo, a line-fitting result independent of that assumption.

\paragraph{H$\alpha$ and H$\beta$ modeling.}
We model the H$\alpha$ and H$\beta$ line profiles of all 11 LRDs with \textsc{rt-scat} for two gas configurations (see Methods): (i)~a \textit{homogeneous} single-phase medium with uniform \NHItwo, and (ii)~a \textit{clumpy} medium with the same total \NHItwo in optically thick clumps embedded in optically thin inter-clump gas, described by a covering factor $f_c$ (the mean number of clumps along a sightline) and a per-clump column $N_{\rm HI,cl} = 3\NHItwo/4f_c$. In both models, the lines are emitted by a central source and scattered by the surrounding $n=2$ \hi gas, after subtraction of the continuum and narrow host emission (see Methods for the source model, host subtraction, and priors). We first fit H$\alpha$ alone and use its \NHItwo posterior as a prior for the H$\beta$ fit, testing whether the H$\alpha$-preferred column also reproduces H$\beta$. 

In the homogeneous model, the fixed H$\alpha$/H$\beta$ scattering cross-section ratio links the two profiles. The H$\alpha$ absorption requires $N_{\rm HI,\mathit{n}=2} \sim 10^{14}$\cmtwo, at which the predicted H$\beta$ absorption is too weak to match the observed depth (Fig.~\ref{fig:example_fit}).
In the clumpy model, two mechanisms break this coupling. First, H$\alpha$ photons scatter off the \textit{surfaces} of optically thick clumps and escape through the inter-clump channels\cite{Neufeld1991,SJC2023,Gronke2016,Gronke2017,Chang2024MgII} without penetrating the high-\NHItwo interiors. Because the competing $3p \to 1s$ (Ly$\beta$) decay is immediately reabsorbed by ground-state hydrogen, H$\alpha$ behaves as a near-perfect resonance line\cite{ChangScattering2026}: photons remain H$\alpha$ across many scatterings, as for Ly$\alpha$, and H$\alpha$ absorption stays weak even at $\NHItwo \sim 10^{16}$\cmtwo. Second, each H$\beta$ scattering has a $\approx$26\% probability of decaying through the $n=3$ level (${\rm BR}_3 \approx 0.26$\cite{ChangScattering2026}), converting the photon into Pa$\alpha$ and H$\alpha$. After $\sim$4 scatterings, $\sim$70\% of H$\beta$ photons are removed, producing strong absorption even where H$\alpha$ escapes freely. The clumpy model therefore reproduces both lines at similar \NHItwo. Extended Data Fig.~\ref{fig:ext_fc} illustrates how both profiles depend on $f_c$ at fixed \NHItwo.

Figure~\ref{fig:example_fit} shows two representative LRDs, GN~9771 and A2744~45924, observed at high and medium NIRSpec resolution. The homogeneous model (blue) cannot reproduce the observed H$\beta$ absorption in either object, whereas the clumpy model (red) fits both lines at $\NHItwo \approx 10^{15.5}$\cmtwo.
Extended Data Figures~\ref{fig:ext_fit1} and \ref{fig:ext_fit2} show the fits for the remaining 9 objects.

\begin{figure*}[ht]
  \centering
  \includegraphics[width=0.94\textwidth]{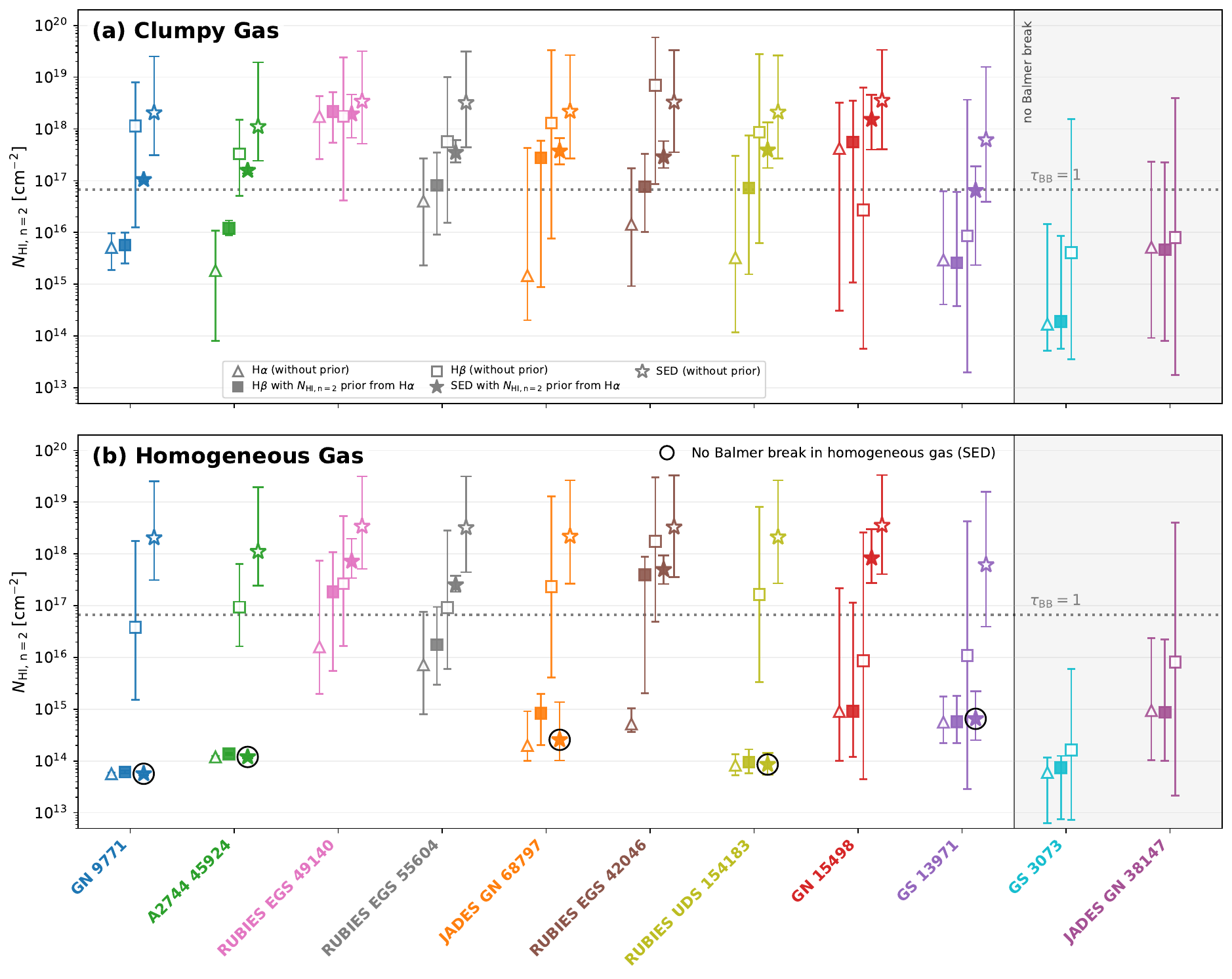}
    \caption{\textbf{$N_{\rm HI,\mathit{n}=2}$ consistency across H$\alpha$,
    H$\beta$, and the Balmer break, for all 11 LRDs.} 
    (a) Clumpy model. (b) Homogeneous model. For each object, triangles show the H$\alpha$ fit; filled squares and filled stars show the H$\beta$ fit and the Balmer-break SED fit, with a \NHItwo prior from the H$\alpha$ fit. Hollow squares and hollow stars show the same H$\beta$ and SED fits without this prior. Error bars are 68 percent credible intervals. The horizontal dotted line marks the column for Balmer-continuum optical depth $\tau_{\rm BB}=1$ ($N_{\rm HI,\mathit{n}=2} \approx 6\times10^{16}$\cmtwo). 
    Black circles in (b) mark homogeneous SED fits that cannot produce the Balmer break (84th percentile of \NHItwo below $10^{16.5}$\cmtwo, i.e. $\tau_{\rm BB}<0.5$): 5 of the 9 break-detected objects. The shaded region marks GS~3073 and JADES~GN~38147, which show no Balmer break. For the clumpy model, \NHItwo of the H$\alpha$ fits and H$\beta$ and SED fits with prior cluster together for most objects, and the SED fit with the prior reaches $\tau_{\rm BB}=1$ for all 9 break-detected objects. For the homogeneous model, \NHItwo from H$\alpha$ is too low to reach the break threshold for 6 of the 9 objects.
}
  \label{fig:nhi_comparison}
\end{figure*}

\paragraph{SED modeling with Balmer break.} 
Under the shared-gas assumption, the Balmer-continuum opacity provides an independent lower bound on \NHItwo. The Balmer-continuum photoionisation cross-section at the series limit is $\sigma_{\rm bc} \approx 1.5\times10^{-17}$\,cm$^2$\,\cite{Karzas1961}, giving unit continuum optical depth at $N_{\rm HI,\mathit{n}=2} \approx 6\times10^{16}$\cmtwo. An observed Balmer break therefore requires $N_{\rm HI,\mathit{n}=2} \gtrsim 10^{17}$\cmtwo.

We fit the SED of each LRD with a three-component model: a power-law UV continuum and an optical blackbody continuum, with a Balmer break and high-$n$ Balmer-series absorption around 3646\,\AA\ (see Methods). 
For 6 of the 9 break-detected objects, the homogeneous \NHItwo from H$\alpha$ lies more than two orders of magnitude below this bound, so the homogeneous model cannot produce the break (Extended Data Figs.~\ref{fig:ext_fit1} and \ref{fig:ext_fit2}; see Methods), whereas the clumpy SED fits with the H$\alpha$ prior reproduce it for all 9.

\paragraph{$N_{\rm HI,\mathit{n}=2}$ tension between Balmer lines and break}
Figure~\ref{fig:nhi_comparison} compares the \NHItwo estimates from H$\alpha$, H$\beta$, and SED. We consider a model to fail for a given object if its \NHItwo from H$\alpha$ lies more than a factor of 100 below the Balmer-break threshold, or the SED fit with the H$\alpha$ prior does not reach it (see Methods and Extended Data Table~\ref{tab:sed_fits}). Of the 9 objects with a detected Balmer break, the clumpy model passes for all 9, and the homogeneous model passes for 3 (RUBIES~EGS~55604, RUBIES~EGS~49140, and GN~15498); for the latter two, the homogeneous H$\alpha$ absorption is at the line center, where its depth depends on the host-line subtraction (see Methods). GS~3073 and JADES~GN~38147 show no Balmer break and are excluded from the pass/fail count (shaded region in Fig.~\ref{fig:nhi_comparison}).  
The H$\beta$ fits with the \NHItwo prior show the same pattern. At \NHItwo from the H$\alpha$ fit, the clumpy model fits H$\beta$ clearly better than the homogeneous model for GN~9771, A2744~45924, and RUBIES~UDS~154183 (see Fig.~\ref{fig:example_fit} and Extended Data Fig.~\ref{fig:ext_fit1}), and the two fit equally well for the other objects. Full \NHItwo posteriors for all three probes, with and without the \NHItwo prior from the H$\alpha$ fit, are shown in Extended Data Fig.~\ref{fig:ext_nhi_post_clumpy} and Fig.~\ref{fig:ext_nhi_post_smooth}.

\begin{figure*}[ht]
  \centering
  \includegraphics[width=\textwidth]{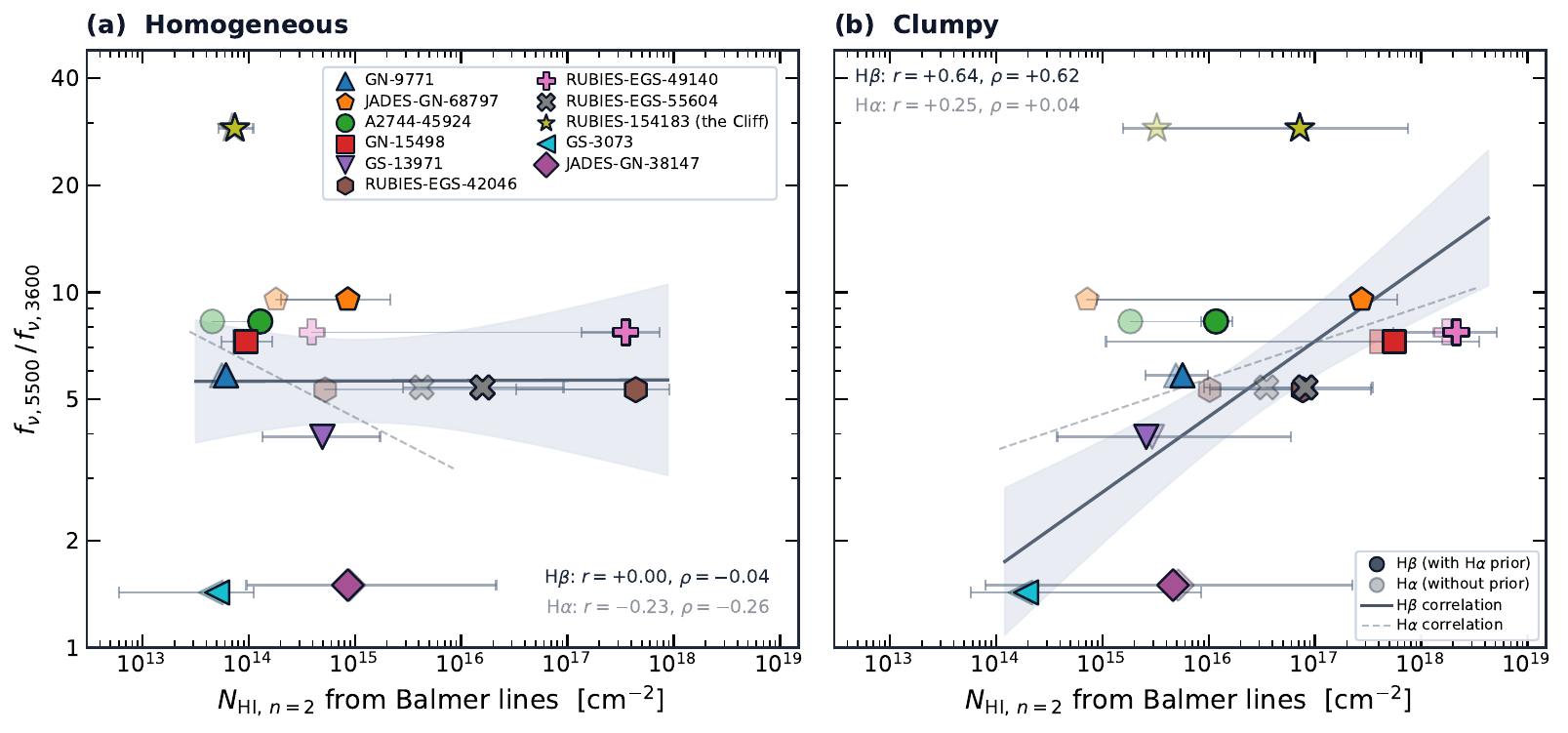}
    \caption{\textbf{Balmer-break color versus $n=2$ neutral hydrogen column density for all 11 LRDs.} The Balmer-break color $f_{\nu,5500}/f_{\nu,3600}$ measures the Balmer-break strength, with larger values indicating a stronger break \cite{Matthee2026}. The left and right panels show it as a function of \NHItwo from the homogeneous and clumpy fits. Each object has a unique marker shape and the same color as in Fig.~\ref{fig:nhi_comparison}. Opaque markers show \NHItwo from the H$\beta$ fit with the prior from the H$\alpha$ fit, and faded markers show the H$\alpha$ fit without the prior. The solid (dashed) lines represent the correlation between $\log(f_{\nu,5500}/f_{\nu,3600})$ versus $\log\,\NHItwo$ from H$\beta$ (H$\alpha$) fit. Pearson ($r$, in log--log space) and Spearman ($\rho$) coefficients are given in each panel. For H$\beta$, the homogeneous model shows no correlation ($r=+0.01$, $\rho=-0.05$), while the clumpy model shows a positive one ($r=+0.64$, $\rho=+0.62$, $p\approx0.04$). For H$\alpha$ alone, neither model shows a significant correlation (clumpy $r=+0.25$, $\rho=+0.04$; homogeneous $r=-0.06$, $\rho=-0.11$); the clumpy slope is positive but not significant.
  }
  \label{fig:nhi_color}
\end{figure*}

\paragraph{\NHItwo and Balmer break strength relation} 

The Balmer-break color, which is the flux ratio at 5500 $\rm \AA$ and 3600 $\rm \AA$, $f_{\nu,5500}/f_{\nu,3600}$, is a direct measure of the break strength \cite{Wang2024,Naidu2025,Matthee2026}.
Figure~\ref{fig:nhi_color} shows the break color as a function of \NHItwo from the H$\beta$ fit with the prior from the H$\alpha$ fit, which is blind to the break, in both clumpy and homogeneous models. We use H$\beta$ because H$\alpha$ alone is only weakly sensitive to \NHItwo in the clumpy model (Methods); the H$\alpha$-only values are shown for comparison. The clumpy model shows a positive correlation with the break strength (Pearson $r=+0.64$, Spearman $\rho=+0.62$), as expected if the clumpy \NHItwo faithfully estimates the true column, whereas the homogeneous model shows no such trend ($r=0.01$, $\rho=-0.05$). This suggests that only the clumpy model links \NHItwo from the Balmer lines to the observed Balmer break.


\paragraph{Relation to current LRD models.} The tension in \NHItwo between the Balmer lines and the Balmer break --- reinforced by the correlation with Balmer-break strength --- shows that a multiphase gas structure (i.e., the clumpy model) is required to explain both features through the same $n=2$ populated \hi gas.
The multiphase gas constraint established here is agnostic to the nature of the central engine and applies across the range of currently proposed LRD models. In the dense gas cocoon scenario \cite{Sneppen2026,Rusakov2025}, the constraint implies that the cocoon must be fragmented on small scales rather than smooth. In the gas-enshrouded black hole and dense accretion disc models \cite{Naidu2025,Chon2026a}, the gas atmosphere must similarly be inhomogeneous to satisfy the joint Balmer-line and Balmer-break constraints.
Black holes can accrete well above the Eddington rate, through radiatively inefficient slim discs or from dense surrounding gas \cite{Abramowicz1988slim,Ohsuga2005,Jiang2014,Inayoshi2016}. In quasi-stars, such accretion proceeds inside a massive envelope \cite{Begelman2008}, a picture since applied to LRDs \cite{Chon2026a,BegelmanDexter2025,Naidu2026}. These models treat the envelope as smooth, although its locally super-Eddington outer layers are expected to be unstable and inhomogeneous \cite{BegelmanDexter2025}. 
If the break instead arises from a supermassive star or an evolved stellar population \cite{Wang2024,Chisholm2026,Chon2026b}, it decouples from the gas, though the Balmer absorption still requires a multiphase $n=2$ \hi medium. 
Our result does not discriminate between these scenarios but identifies a necessary ingredient, strong small-scale inhomogeneity in the $n=2$ H~I gas, that any viable model must accommodate. The broad line wings do not enter this test. Whether they arise from Thomson scattering in an inner ionized region \cite{ChangScattering2026} or the broad-line region itself \cite{MadauMaiolino2026}, in our model they only set the incident line profile; the Balmer absorption and break form in the $n=2$ \hi gas outside the emission region, which must be multiphase in either case.
The H$\alpha$ and H$\beta$ profiles alone also require multiphase structure, independently of whether the break arises from the same gas.

\paragraph{Multiphase gas structure and formation} 
The clumpy geometry required by our modeling can be interpreted physically as a multiphase medium: dense, cool, $n=2$ enhanced H~I clumps with covering factor $f_c\sim5$--$60$ and individual column densities $N_{\rm HI,cl}\sim10^{15}$--$10^{17}$\cmtwo, coexisting with a diffuse, optically thin interclump medium (see Extended Data Fig.~\ref{fig:ext_geometry}). The total column density \NHItwo and $f_c$ are constrained by the Balmer-line modeling, but the large-scale morphology, whether a discrete cloud ensemble, porous shell, or fragmented disc, is not determined by the present data alone. 
A clumpy neutral medium has also been considered a viable explanation for the Ly$\alpha$ profile of the lensed LRD QSO1 \cite{Tang2026}.

Such multiphase structure is common around AGNs. Broad emission lines are attributed to small, dense ($n_{\rm H}\sim10^{10}\,\rm cm^{-3}$) clouds with a volume filling fraction of $\sim10^{-6}$, confined by a hotter medium or magnetic fields \cite{Rees1987,Netzer2006}; smooth line profiles require $\gtrsim10^{6}$--$10^{8}$ such emitters \cite{Arav1997,Dietrich1999}, and internal turbulence of $\sim10^{3}$\kms is allowed in the line-emitting gas \cite{Bottorff2000}. Outflowing absorbers, from broad and mini broad absorption line regions near the accretion disc \cite{Elvis2000,Hamann2011,Hamann2013} to kpc-scale FeLoBAL and associated absorbers \cite{FaucherGiguere2012,DOdorico2004,Finn2014}, likewise resolve into multiple narrow components. Our fits are consistent with this picture: for $N_{\rm cl}$ clumps of radius $r_{\rm cl}$ within a region of size $R$, $f_c\sim N_{\rm cl}(r_{\rm cl}/R)^{2}$, so $f_c\sim5$--$60$ implies $N_{\rm cl}\sim10^{5}$--$10^{6}$ for $r_{\rm cl}/R\sim10^{-2}$, and more for smaller clumps\footnote{The radiative transfer depends on $f_c$ and $N_{\rm HI,cl}$ but not on clump size for small volume filling factors \cite{Hansen2006,SJC2023}, so $r_{\rm cl}$ and $N_{\rm cl}$ are not individually constrained.}, comparable to BLR and BAL cloud numbers. However, BLR clouds are photoionized and emit the broad lines, whereas the clumps in our model are a neutral scattering medium with a substantial $n=2$ population.

Such clumps can form and survive in this environment. Turbulent multiphase gas fragments into many small structures rather than remaining monolithic, and mixing at the interfaces with the hot gas induces rapid cooling, so the cool phase grows rather than being destroyed \cite{Gronke2022,GronkeSchneider2026}.
Since only a small fraction of the hydrogen is in $n=2$, the total clump columns implied by $N_{\rm HI,cl}\sim10^{15}$--$10^{17}$\cmtwo exceed both the fragmentation scale of rapidly cooling gas ($\sim10^{17}$\cmtwo\cite{McCourt2018}) and the minimum column for survival in a hot wind ($\sim10^{18}$\cmtwo\cite{GronkeOh2018}).
The clump population implied by our fits is thus a natural outcome of cooling in a turbulent medium, although the data do not constrain its formation history.

Cool clumps alone do not explain their large $n=2$ population. Multiple scatterings of Ly$\alpha$ in optically thick clumps can pump $n=2$ near their surfaces \cite{Dijkstra2016a,Dijkstra2016b,Seon2020}, and collisional excitation may contribute in dense gas; comparing the observed Ly$\alpha$ and H$\alpha$ fluxes can constrain their relative importance \cite{Tang2026}, and a detailed calculation will be presented in a companion paper.

One circumnuclear setting for such gas is a thermally unstable disc wind, in which cool clumps condense from a hot outflow. The densities invoked in previous LRD studies, $n_{\rm H}\sim10^{9}$--$10^{11}$\,cm$^{-3}$ \cite{Naidu2025,Inayoshi2025}, overlap with those of broad-line region clouds ($n_{\rm H}\sim10^{8}$--$10^{10}$\,cm$^{-3}$), which are commonly associated with clumpy disc winds \cite{Murray1995,Elvis2000}.
We therefore suggest, cautiously, that the multiphase $n=2$ \hi clumps in LRDs may represent an early BLR-like or proto-BLR medium formed during rapid black hole growth. In this picture, what distinguishes LRDs from ordinary broad-line AGNs is not the presence of clumps but their ionization state. Shielded by the dense inner gas of a super-Eddington flow, the clumps stay neutral. As the accretion rate decreases and the disc thins \cite{Abramowicz1988slim,Inayoshi2025b,Chon2026a}, reduced shielding could induce ionization of these clumps, eventually giving rise to a standard BLR, which would also explain why LRD-like spectra are rare at lower redshift \cite{Kocevski2025,Lin2025}. The present data establish the presence of strongly inhomogeneous $n=2$ \hi gas, but do not uniquely determine its disc-wind origin or identify the clumps as proto-BLR clouds.

\paragraph{Origin of the red optical continuum}

The multiphase gas constraint does not by itself determine the origin of the red optical continuum in LRDs. Proposed origins include a super-Eddington photosphere \cite{Liu2025,Naidu2025}, supermassive stars \cite{Chisholm2026,Chon2026b}, a transmitted AGN SED \cite{Inayoshi2025,Killi2024}, and nebular continuum emission \cite{Sneppen2026,Ando2026}. The nebular case is directly testable, because there the continuum and the hydrogen lines both arise from recombination, whereas the other models place the continuum in a region separate from the line emission. Since H$\alpha$ and the nebular continuum both scale with the emission measure, a pure-recombination spectrum from low-density gas (where collisions are negligible) that is optically thin in the Balmer lines has an intrinsic H$\alpha$ equivalent width (EW) of ${\sim}5{,}000$--$10{,}000$\,\AA\ regardless of geometry (Extended Data Fig.~\ref{fig:ext_neb_bb_ew}), up to $10$--$100$ times larger than the observed LRD values. Reproducing the observed continuum from nebular gas alone therefore requires either substantial H$\alpha$ suppression, relaxing one of these two assumptions, or a separate, non-nebular continuum component.

Relaxing the optically thin assumption, bound--free absorption from $n=3$ has been suggested to suppress H$\alpha$ \cite{Sneppen2026}. Resonance scattering lengthens the $n=3$ column an H$\alpha$ photon samples relative to the continuum, but our RT calculations give a path-length enhancement of only ${\sim}2$--$30$ (Extended Data Fig.~\ref{fig:ext_path_length})\footnote{Dust along this path could likewise extinguish H$\alpha$ relative to the continuum, though direct evidence for dust in this gas is lacking (see Methods).}, two to four orders of magnitude short of what is required (see Methods). 
The $n=3$ column needed instead predicts near-saturated Paschen-line absorption, which is not seen in LRDs with Paschen-band observations and clear Balmer absorption \cite{Brazzini2025,Kokorev2026GLIMPSE, Kiyota2026}.

In denser gas, collisional de-excitation of $n=3$ offers a second route. Because a trapped H$\alpha$ photon revisits $n=3$ repeatedly before escaping, this channel becomes accessible at $n_e\sim10^{9}$--$10^{11}$\,cm$^{-3}$, within the range invoked in previous LRD studies \cite{Naidu2025,Inayoshi2025}, although the observed H$\beta$ survival caps how optically thick the emission region can be \cite{ChangScattering2026} (see Methods). Both channels predict the same Paschen-line signature, so we cannot determine which, if either, resolves the puzzle. Either way, the EW argument requires H$\alpha$ suppression or a continuum source distinct from the line-emitting gas, regardless of whether the $n=2$ medium is clumpy.

\paragraph{Caveats, robustness, and observational predictions} 
The line--break consistency test assumes that the same $n=2$ \hi gas produces both the Balmer-line absorption and the Balmer break. If the break instead formed in a separate continuum-emitting region, as in photospheric or stellar models of the red continuum, the line--break comparison alone would not require clumpiness. That region would, however, need an \NHItwo that happens to track the Balmer-line absorption, which the suggestive correlation between the H$\beta$-fit \NHItwo and break strength disfavors but cannot exclude.
The line-profile fits alone do not rely on this assumption: at \NHItwo from the H$\alpha$ fit, the homogeneous model cannot reproduce the observed H$\beta$ absorption in several objects, whereas the clumpy model reproduces both lines.

Host-galaxy subtraction, the unresolved H$\beta$ core at $\mathcal{R}\sim1000$, and an \NHItwo--$f_c$ degeneracy when no clear absorption is detected affect the exact parameters of individual objects (see Methods), 
but do not explain why the homogeneous model fails to fit both lines while the clumpy model succeeds.

Our radiative transfer treats \NHItwo as a free parameter and neglects collisional processes. In practice, the $n=2$ and $n=3$ populations are set jointly by Ly$\alpha$ pumping, collisional excitation and de-excitation, and line trapping, all of which depend on gas density; coupling these to the radiative transfer would translate the fitted \NHItwo and $f_c$ into physical densities and clump sizes. 

The multiphase model makes several testable predictions. Higher Balmer lines share the branching loss of H$\beta$: excited $np$ states with $n \geq 4$ can decay through intermediate levels $2<n'<n$, emitting Paschen, Brackett, or higher-series photons instead of the original Balmer photon. H$\gamma$ and H$\delta$ should therefore show absorption comparable to H$\beta$, not intermediate between H$\alpha$ and H$\beta$: the transition is a step at H$\alpha$, the only near-perfect resonance line, rather than a gradual hierarchy. 
Saturated H$\gamma$ absorption has already been invoked for the broad H$\gamma$ profile of GN~9771 \cite{Torralba2026}, and deep high-resolution spectroscopy now resolves H$\alpha$--H$\delta$ absorption in individual LRDs \cite{Wang2026}; a test across the sample awaits higher signal-to-noise, as these lines are intrinsically weaker.
For the LRDs whose H$\beta$ cores remain unresolved or have insufficient signal-to-noise, high-resolution NIRSpec follow-up can test whether they show the same H$\alpha$--H$\beta$ discrepancy already unambiguous in GN~9771 ($\mathcal{R} = 2700$). 
Finally, Paschen-band spectroscopy would test both H$\alpha$ suppression channels discussed above at once: near-saturated Paschen-line absorption is expected if either operates, whereas its absence would favor a non-nebular contribution to the red optical continuum.

\backmatter

\bmhead{Acknowledgements}
This work is based on observations made with the NASA/ESA/CSA James Webb Space Telescope. The data were obtained from the Mikulski Archive for Space Telescopes (MAST) at the Space Telescope Science Institute, which is operated by the Association of Universities for Research in Astronomy, Inc., under NASA contract NAS 5-03127 for JWST. These observations are associated with programs \#1181, 1216, 4233, 5664, 8204 and 9433. Support for program \#5664 was provided by NASA through a grant from the Space Telescope Science Institute. The authors acknowledge the RUBIES team, led by co-PIs A. de Graaff and G. Brammer, and the \#9433 team, led by co-PIs R. Maiolino and F. D'Eugenio, for developing their observing programs with a zero-exclusive-access period.
SJC acknowledges support from the ERC synergy grant 101166930 - RECAP. 
JM acknowledges funding by the European Union (ERC, AGENTS,  101076224).
MG thanks the European Union for support through ERC-2024-STG 101165038 (ReMMU).
CAM acknowledges support by the European Union ERC grant RISES (101163035), Carlsberg Foundation (CF22-1322), and VILLUM FONDEN (37459). The Cosmic Dawn Center (DAWN) is funded by the Danish National Research Foundation under grant DNRF140.
The authors used an AI assistant (Claude, Anthropic) for building the fitting pipeline and language editing. All analyses, interpretations, and conclusions were made and verified by the authors, who take full responsibility for the content.

\section*{Declarations}

\begin{itemize}
\item \textbf{Funding}: S.-J.C. discloses support for this work from the ERC synergy grant 101166930 (RECAP). J.M. discloses support from the European Union (ERC, AGENTS, 101076224). M.G. discloses support from the European Union (ERC-2024-STG 101165038, ReMMU). C.M. discloses support from the European Union ERC grant RISES (101163035), the Carlsberg Foundation (CF22-1322), and VILLUM FONDEN (37459); the Cosmic Dawn Center (DAWN) is funded by the Danish National Research Foundation under grant DNRF140.
\item \textbf{Competing interests}: The authors declare no competing interests.
\item \textbf{Ethics approval and consent to participate}: Not applicable.
\item \textbf{Consent for publication}: Not applicable.
\item \textbf{Data availability}: The JWST NIRSpec spectra used in this work are described in \cite{Matthee2026}. This work is based on observations made with the NASA/ESA/CSA James Webb Space Telescope, obtained from the Mikulski Archive for Space Telescopes at the Space Telescope Science Institute, operated by the Association of Universities for Research in Astronomy, Inc., under NASA contract NAS 5-03127 for JWST. These observations are associated with programs \#1181, 1216, 4233, 5664, 8204 and 9433 (see Extended Data Table~\ref{tab:sample}). The radiative-transfer model grids and MCMC posterior chains are available from the corresponding author upon reasonable request and will be deposited in a public repository upon acceptance (DOI to be confirmed upon acceptance).  
\item \textbf{Materials availability}: Not applicable.
\item \textbf{Code availability}: The \textsc{rt-scat} radiative-transfer code is described in \cite{Chang2024MgII,ChangScattering2026}. Posterior sampling uses the affine-invariant ensemble sampler \textsc{emcee} \cite{ForemanMackey2013}. The MCMC fitting pipeline built around these tools is available from the corresponding author upon reasonable request.
\item \textbf{Author contributions}:
S.-J.C. led the project, performed the radiative-transfer simulations, analyzed the simulated data, and modeled the Balmer lines and SED.
M.G. contributed to building the fitting pipeline and to the scientific interpretation of the fitting results and their connection to the multiphase nature of the gas.
J.M. provided the observed spectra for the sample of 11 LRDs, contributed to the interpretation of the data and to connecting the modeling results to LRD observations, and offered insight into how the results compare with other proposed LRD models.
C.M. contributed to the observational interpretation of early-universe and LRD observations and to checking the models against recent observational results.
\end{itemize}

\bibliography{references}

@ARTICLE{Killi2024,
       author = {{Killi}, Meghana and {Watson}, Darach and {Brammer}, Gabriel and {McPartland}, Conor and {Antwi-Danso}, Jacqueline and {Newshore}, Rosa and {Coe}, Dan and {Allen}, Natalie and {Fynbo}, Johan P.~U. and {Gould}, Katriona and {Heintz}, Kasper E. and {Rusakov}, Vadim and {Vejlgaard}, Simone},
        title = "{Deciphering the JWST spectrum of a 'little red dot' at z {\ensuremath{\sim}} 4.53: An obscured AGN and its star-forming host}",
      journal = {\aap},
         year = 2024,
        month = nov,
       volume = {691},
          eid = {A52},
        pages = {A52},
          doi = {10.1051/0004-6361/202348857},
archivePrefix = {arXiv},
       eprint = {2312.03065},
 primaryClass = {astro-ph.GA},
       adsurl = {https://ui.adsabs.harvard.edu/abs/2024A&A...691A..52K}
}

@ARTICLE{Lin2025,
       author = {{Lin}, Ruqiu and {Zheng}, Zhen-Ya and {Jiang}, Chunyan and {Yuan}, Fang-Ting and {Ho}, Luis C. and {Wang}, Junxian and {Jiang}, Linhua and {Rhoads}, James E. and {Malhotra}, Sangeeta and {Barrientos}, L. Felipe and {Wold}, Isak and {Infante}, Leopoldo and {Zhu}, Shuairu and {Ji}, Xiang and {Fu}, Xiaodan},
        title = "{Discovery of Local Analogs to JWST's Little Red Dots}",
      journal = {\apjl},
         year = 2025,
        month = feb,
       volume = {980},
       number = {2},
          eid = {L34},
        pages = {L34},
          doi = {10.3847/2041-8213/adaaf1},
archivePrefix = {arXiv},
       eprint = {2412.08396},
 primaryClass = {astro-ph.GA},
       adsurl = {https://ui.adsabs.harvard.edu/abs/2025ApJ...980L..34L}
}

@ARTICLE{Kocevski2025,
       author = {{Kocevski}, Dale D. and {Finkelstein}, Steven L. and {Barro}, Guillermo and {Taylor}, Anthony J. and {Calabr{\`o}}, Antonello and {Laloux}, Brivael and {Buchner}, Johannes and {Trump}, Jonathan R. and {Leung}, Gene C.~K. and {Yang}, Guang and {Dickinson}, Mark and {P{\'e}rez-Gonz{\'a}lez}, Pablo G. and {Pacucci}, Fabio and {Inayoshi}, Kohei and {Somerville}, Rachel S. and {McGrath}, Elizabeth J. and {Akins}, Hollis B. and {Bagley}, Micaela B. and {Bowler}, Rebecca A.~A. and {Bisigello}, Laura and {Carnall}, Adam and {Casey}, Caitlin M. and {Cheng}, Yingjie and {Cleri}, Nikko J. and {Costantin}, Luca and {Cullen}, Fergus and {Davis}, Kelcey and {Donnan}, Callum T. and {Dunlop}, James S. and {Ellis}, Richard S. and {Ferguson}, Henry C. and {Fujimoto}, Seiji and {Fontana}, Adriano and {Giavalisco}, Mauro and {Grazian}, Andrea and {Grogin}, Norman A. and {Hathi}, Nimish P. and {Hirschmann}, Michaela and {Huertas-Company}, Marc and {Holwerda}, Benne W. and {Illingworth}, Garth and {Juneau}, St{\'e}phanie and {Kartaltepe}, Jeyhan S. and {Koekemoer}, Anton M. and {Li}, Wenxiu and {Lucas}, Ray A. and {Magee}, Dan and {Mason}, Charlotte and {McLeod}, Derek J. and {McLure}, Ross J. and {Napolitano}, Lorenzo and {Papovich}, Casey and {Pirzkal}, Nor and {Rodighiero}, Giulia and {Santini}, Paola and {Wilkins}, Stephen M. and {Yung}, L.~Y. Aaron},
        title = "{The Rise of Faint, Red Active Galactic Nuclei at z > 4: A Sample of Little Red Dots in the JWST Extragalactic Legacy Fields}",
      journal = {\apj},
         year = 2025,
        month = jun,
       volume = {986},
       number = {2},
          eid = {126},
        pages = {126},
          doi = {10.3847/1538-4357/adbc7d},
archivePrefix = {arXiv},
       eprint = {2404.03576},
 primaryClass = {astro-ph.GA},
       adsurl = {https://ui.adsabs.harvard.edu/abs/2025ApJ...986..126K}
}

@ARTICLE{Kiyota2026,
       author = {{Kiyota}, Tomokazu and {Ouchi}, Masami and {Yanagisawa}, Hiroto and {Ando}, Makoto and {Harikane}, Yuichi and {Kageura}, Yuta and {Nakane}, Minami and {Nakazato}, Yurina and {Ono}, Yoshiaki and {Takeda}, Yui},
        title = "{ATLAS. III. Dust Around Little Red Dots: Hydrogen Line Ratios beyond Dust-free Non-Case B Models}",
      journal = {arXiv e-prints},
         year = 2026,
        month = aug,
          eid = {arXiv:2608.10832},
        pages = {arXiv:2608.10832},
          doi = {10.48550/arXiv.2608.10832},
archivePrefix = {arXiv},
       eprint = {2608.10832},
 primaryClass = {astro-ph.GA},
       adsurl = {https://ui.adsabs.harvard.edu/abs/2026arXiv260810832K}
}

@ARTICLE{Davis2026,
       author = {{Davis}, Kelcey and {Brooks}, Madisyn and {Simons}, Raymond C. and {Trump}, Jonathan R. and {Barro}, Guillermo and {Arrabal Haro}, Pablo and {Backhaus}, Bren E. and {Cleri}, Nikko J. and {de la Vega}, Alexander and {Finkelstein}, Steven L. and {Giavalisco}, Mauro and {Grogin}, Norman A. and {Hirschmann}, Michaela and {Hutchison}, Taylor A. and {Kocevski}, Dale and {Koekemoer}, Anton M. and {Lambrides}, Erini and {Llerena}, Mario and {Lucas}, Ray A. and {Marshall}, Madeline A. and {McGrath}, Elizabeth J. and {Papovich}, Casey and {Starrs}, Aidan and {Taylor}, Anthony J. and {Upton Sanderbeck}, Phoebe R. and {Wang}, Xin and {Wuyts}, Stijn},
        title = "{OCEANS of Absorption: High-resolution NIRSpec Spectroscopy Reveals Diverse Balmer-line Absorption in Little Red Dots}",
      journal = {arXiv e-prints},
         year = 2026,
        month = may,
          eid = {arXiv:2606.00258},
        pages = {arXiv:2606.00258},
          doi = {10.48550/arXiv.2606.00258},
archivePrefix = {arXiv},
       eprint = {2606.00258},
 primaryClass = {astro-ph.GA},
       adsurl = {https://ui.adsabs.harvard.edu/abs/2026arXiv260600258D}
}

@ARTICLE{Juodzbalis2026,
       author = {{Juod{\v{z}}balis}, Ignas and {Ji}, Xihan and {D'Eugenio}, Francesco and {Scholtz}, Jan and {Maiolino}, Roberto and {Stoffers}, Amanda and {Marconi}, Alessandro and {Bertola}, Elena and {Bunker}, Andrew J. and {Carniani}, Stefano and {Cresci}, Giovanni and {Curtis-Lake}, Emma and {Ma}, Zheng and {Marconcini}, Cosimo and {Parlanti}, Eleonora and {Rinaldi}, Pierluigi and {Robertson}, Brant and {{\"U}bler}, Hannah and {Venturi}, Giacomo and {Zhang}, Junyu},
        title = "{An (in)complete NIRSpec census of Balmer absorption in Type 1 AGN -- radiation-driven outflows in little red dots, quasars and variable stars}",
      journal = {arXiv e-prints},
         year = 2026,
        month = jul,
          eid = {arXiv:2607.19057},
        pages = {arXiv:2607.19057},
          doi = {10.48550/arXiv.2607.19057},
archivePrefix = {arXiv},
       eprint = {2607.19057},
 primaryClass = {astro-ph.GA},
       adsurl = {https://ui.adsabs.harvard.edu/abs/2026arXiv260719057J}
}

@ARTICLE{Wang2026,
  author = {{Wang}, Bingjie and {Greene}, Jenny E. and {Liu}, Hanpu and {Kaaz}, Nicholas and {Brammer}, Gabriel B. and 
  others},
  title = "{Balmer Absorption Series and Broad Metal Lines in Two Luminous Little Red Dots}",
  journal = {arXiv e-prints},
  year = 2026,
  month = sep,
  eid = {arXiv:2609.22623},
  eprint = {2609.22623},
  archivePrefix = {arXiv},
  primaryClass = {astro-ph.GA}
  }

@ARTICLE{Cooper2026,
       author = {{Cooper}, Ryan A. and {Caputi}, Karina I. and {Marconi}, Alessandro and {Feltre}, Anna and {Iani}, Edoardo and {Rinaldi}, Pierluigi and {Desprez}, Guillaume and {Katz}, Harley},
        title = "{The Physical Conditions of Low-Mass Galaxies at z = 3.5 - 7.0 from JWST Spectroscopy: The Behaviour of Spectral Line Ratios with Burstiness}",
      journal = {arXiv e-prints},
         year = 2026,
        pages = {arXiv:2609.01205},
          doi = {10.48550/arXiv.2609.01205},
       eprint = {2609.01205}
}

@ARTICLE{Inayoshi2016,
       author = {{Inayoshi}, Kohei and {Haiman}, Zolt{\'a}n and {Ostriker}, Jeremiah P.},
        title = "{Hyper-Eddington accretion flows on to massive black holes}",
      journal = {MNRAS},
       volume = {459},
         year = 2016,
        pages = {3738},
          doi = {10.1093/mnras/stw836}
}

@ARTICLE{Begelman2008,
       author = {{Begelman}, Mitchell C. and {Rossi}, Elena M. and {Armitage}, Philip J.},
        title = "{Quasi-stars: accreting black holes inside massive envelopes}",
      journal = {MNRAS},
       volume = {387},
         year = 2008,
        pages = {1649},
          doi = {10.1111/j.1365-2966.2008.13344.x}
}

@ARTICLE{BegelmanDexter2025,
       author = {{Begelman}, Mitchell C. and {Dexter}, Jason},
        title = "{Little Red Dots as Late-stage Quasi-stars}",
      journal = {ApJ},
       volume = {996},
         year = 2025,
        pages = {48},
          doi = {10.3847/1538-4357/ae274a}
}

@ARTICLE{MadauMaiolino2026,
       author = {{Madau}, Piero and {Maiolino}, Roberto},
        title = "{Little red dots as obscured little blue dots: A super-Eddington unification model}",
      journal = {A\&A},
         year = 2026,
          doi = {10.1051/0004-6361/202659606},
       eprint = {2602.22386}
}

@ARTICLE{Hansen2006,
       author = {{Hansen}, Mark and {Oh}, S. Peng},
        title = "{Lyman $\alpha$ radiative transfer in a multiphase medium}",
      journal = {MNRAS},
       volume = {367},
         year = 2006,
        pages = {979},
          doi = {10.1111/j.1365-2966.2005.09870.x}
}

@ARTICLE{Arav1997,                                                                                                                            
       author = {{Arav}, N. and {Barlow}, T. A. and {Laor}, A. and {Blandford}, R. D.},
        title = "{Keck high-resolution spectroscopy of Mrk 335: constraints on the number of emitting clouds in the broad-line region}",
      journal = {MNRAS},
       volume = {288},
         year = 1997,
        pages = {1015},
          doi = {10.1093/mnras/288.4.1015}
}

@ARTICLE{Bottorff2000,
       author = {{Bottorff}, Mark and {Ferland}, Gary and {Baldwin}, Jack and {Korista}, Kirk},
        title = "{Observational Constraints on the Internal Velocity Field of Quasar Emission-Line Clouds}",
      journal = {ApJ},
       volume = {542},
         year = 2000,
        pages = {644},
          doi = {10.1086/317051}
}

@ARTICLE{DOdorico2004,
       author = {{D'Odorico}, Valentina and {Cristiani}, Stefano and {Romano}, Donatella and {Granato}, Gian Luigi and {Danese}, Luigi},
        title = "{Chemical abundances in QSO host galaxies and environments from narrow absorption line systems}",
      journal = {MNRAS},
       volume = {351},
         year = 2004,
        pages = {976},
          doi = {10.1111/j.1365-2966.2004.07840.x}
}

@ARTICLE{FaucherGiguere2012,
       author = {{Faucher-Gigu{\`e}re}, Claude-Andr{\'e} and {Quataert}, Eliot and {Murray}, Norman},
        title = "{A physical model of FeLoBALs: implications for quasar feedback}",
      journal = {MNRAS},
       volume = {420},
         year = 2012,
        pages = {1347},
          doi = {10.1111/j.1365-2966.2011.20120.x}
}

@ARTICLE{Finn2014,
       author = {{Finn}, Charles W. and {Morris}, Simon L. and {Crighton}, Neil H. M. and {Hamann}, Fred and {Done}, Chris and {Theuns}, Tom and {Fumagalli}, Michele and {Tejos}, Nicolas and {Worseck}, Gabor},
        title = "{A compact, metal-rich, kpc-scale outflow in FBQS J0209$-$0438: detailed diagnostics from HST/COS extreme UV observations}",
      journal = {MNRAS},
       volume = {440},
         year = 2014,
        pages = {3317},
          doi = {10.1093/mnras/stu518}
}

@ARTICLE{GronkeOh2018,
       author = {{Gronke}, Max and {Oh}, S. Peng},
        title = "{The growth and entrainment of cold gas in a hot wind}",
      journal = {MNRAS},
       volume = {480},
         year = 2018,
        pages = {L111},
          doi = {10.1093/mnrasl/sly131}
}

@ARTICLE{Hamann2011,
       author = {{Hamann}, F. and {Kanekar}, N. and {Prochaska}, J. X. and {Murphy}, M. T. and {Ellison}, S. and {Malec}, A. L. and {Milutinovic}, N. and {Ubachs}, W.},
        title = "{A high-velocity narrow absorption line outflow in the quasar J212329.46$-$005052.9}",
      journal = {MNRAS},
       volume = {410},
         year = 2011,
        pages = {1957},
          doi = {10.1111/j.1365-2966.2010.17575.x}
}

@ARTICLE{Hamann2013,
       author = {{Hamann}, F. and {Chartas}, G. and {McGraw}, S. and {Rodriguez Hidalgo}, P. and {Shields}, J. and {Capellupo}, D. and {Charlton}, J. and {Eracleous}, M.},
        title = "{Extreme-velocity quasar outflows and the role of X-ray shielding}",
      journal = {MNRAS},
       volume = {435},
         year = 2013,
        pages = {133},
          doi = {10.1093/mnras/stt1231}
}

@ARTICLE{Rees1987,
       author = {{Rees}, M. J.},
        title = "{Magnetic confinement of broad-line clouds in active galactic nuclei}",
      journal = {MNRAS},
       volume = {228},
         year = 1987,
        pages = {47P},
          doi = {10.1093/mnras/228.1.47p}
}

@ARTICLE{Dietrich1999,
       author = {{Dietrich}, M. and {Wagner}, S. J. and {Courvoisier}, T. J. -L. and {Bock}, H. and {North}, P.},
        title = "{Structure of the broad-line region of 3C 273}",
      journal = {A\&A},
       volume = {351},
         year = 1999,
        pages = {31}
}

@INCOLLECTION{Netzer2006,
       author = {{Netzer}, Hagai},
        title = "{Active Galactic Nuclei: Basic Physics and Main Components}",
    booktitle = {Physics of Active Galactic Nuclei at all Scales},
       editor = {{Alloin}, Danielle and {Johnson}, Rachel and {Lira}, Paulina},
       series = {Lecture Notes in Physics},
       volume = {693},
    publisher = {Springer},
         year = 2006,
        pages = {1},
          doi = {10.1007/3-540-34621-X_1}
}

@ARTICLE{Brazzini2025,
       author = {{Brazzini}, M. and {D'Eugenio}, F. and {Maiolino}, R. and {Juodzbalis}, I. and {Ji}, X. and {Scholtz}, J. and {Chang}, S.-J.},
        title = "{Ruling out dominant electron scattering in Little Red Dots' Rosetta Stone using multiple hydrogen lines}", 
      journal = {MNRAS},
       volume = {544},
         year = 2025,
        pages = {L167},
          doi = {10.1093/mnrasl/slaf116}
}

@ARTICLE{Chang2018,
       author = {{Chang}, Seok-Jun and {Lee}, Hee-Won and {Ahn}, Sang-Hyeon and {Lee}, Hogyu and {Angeloni}, Rodolfo and {Palma}, Tali and {Di Mille}, Francesco},
        title = "{Escape of Resonantly Scattered Ly$\beta$ and H$\alpha$ from Hot and Optically Thick Media}",
      journal = {JKAS},
       volume = {51},
         year = 2018,
        pages = {5},
          doi = {10.5303/JKAS.2018.51.1.5}
}

@ARTICLE{Kido2025,                                                                                                                
       author = {{Kido}, Daisaburo and {Ioka}, Kunihito and {Hotokezaka}, Kenta and {Inayoshi}, Kohei and {Irwin}, Christopher M.},
        title = "{Black hole envelopes in Little Red Dots}",
      journal = {MNRAS},
       volume = {544},
         year = 2025,
        pages = {3407},
          doi = {10.1093/mnras/staf1898}
}

@ARTICLE{Hviding2025,
       author = {{Hviding}, Raphael E. and {de Graaff}, Anna and {Miller}, Tim B. and {Setton}, David J. and {Greene}, Jenny E. and {Labb{\'e}}, Ivo and {Brammer}, Gabriel and {Bezanson}, Rachel and {Boogaard}, Leindert A. and {Cleri}, Nikko J. and {Leja}, Joel and {Maseda}, Michael V. and {McConachie}, Ian and {Matthee}, Jorryt and {Naidu}, Rohan P. and {Oesch}, Pascal A. and {Wang}, Bingjie and {Whitaker}, Katherine E. and {Williams}, Christina C.},
        title = "{RUBIES: A spectroscopic census of little red dots. All point sources with v-shaped continua have broad lines}",
      journal = {A\&A},
       volume = {702},
         year = 2025,
        pages = {A57},
          doi = {10.1051/0004-6361/202555816}
}

@ARTICLE{dEugenio2026,
       author = {{D'Eugenio}, Francesco and {Maiolino}, Roberto and {Perna}, Michele and {{\"U}bler}, Hannah and {Ji}, Xihan and others},
        title = "{BlackTHUNDER strikes twice: Balmer-line absorption in an overmassive Little Red Dot at $z = 7.04$}",
      journal = {MNRAS},
       volume = {547},
         year = 2026,
        pages = {1},
          doi = {10.1093/mnras/stag401}
}

@ARTICLE{Ando2026,
       author = {{Ando}, Makoto and {Harikane}, Yuichi and {Katz}, Harley and {Inayoshi}, Kohei and {Tanaka}, Takumi S.},
        title = "{The UV Side of Little Red Dots: Red, Compact, and Iron-Enhanced Rest-UV Emission with a Strong Downturn around Ly{\ensuremath{\alpha}}}",
      journal = {arXiv e-prints},
         year = 2026,
        pages = {arXiv:2606.03522},
          doi = {10.48550/arXiv.2606.03522},
       eprint = {2606.03522}
}

@ARTICLE{deGraaff2025,
       author = {{de Graaff}, Anna and {Rix}, Hans-Walter and {Naidu}, Rohan P. and {Labbe}, Ivo and {Wang}, Bingjie and {Leja}, Joel and {Matthee}, Jorryt and {Katz}, Harley and {Greene}, Jenny E. and {Hviding}, Raphael E. and others},
        title = "{A remarkable ruby: Absorption in dense gas, rather than evolved stars, drives the extreme Balmer break of a little red dot at z = 3.5}",
      journal = {Astronomy \& Astrophysics},
         year = 2025,
       volume = {701},
        pages = {A168},
          doi = {10.1051/0004-6361/202554681}
}

@ARTICLE{Abramowicz1988slim,
       author = {{Abramowicz}, M.~A. and {Czerny}, B. and {Lasota}, J.~P. and {Szuszkiewicz}, E.},
        title = "{Slim Accretion Disks}",
      journal = {Astrophysical Journal},
         year = 1988,
       volume = {332},
        pages = {646--658},
          doi = {10.1086/166683}
}

@ARTICLE{GronkeSchneider2026,
       author = {{Gronke}, Max and {Schneider}, Evan E.},
        title = "{Simulations of multi-phase gas in and around galaxies}",
      journal = {Living Reviews in Computational Astrophysics},
         year = 2026,
       volume = {12},
        pages = {2},
          doi = {10.1007/s41115-026-00025-7},
       eprint = {2601.16566}
}

@ARTICLE{Gronke2022,
       author = {{Gronke}, Max and {Oh}, S.~Peng and {Ji}, Suoqing and {Norman}, Colin},
        title = "{Survival and mass growth of cold gas in a turbulent, multiphase medium}",
      journal = {Monthly Notices of the Royal Astronomical Society},
         year = 2022,
       volume = {511},
        pages = {859--876},
          doi = {10.1093/mnras/stab3351},
       eprint = {2107.13012}
}

@ARTICLE{McCourt2018,
       author = {{McCourt}, Michael and {Oh}, S.~Peng and {O'Leary}, Ryan and {Madigan}, Ann-Marie},
        title = "{A characteristic scale for cold gas}",
      journal = {Monthly Notices of the Royal Astronomical Society},
         year = 2018,
       volume = {473},
        pages = {5407--5431},
          doi = {10.1093/mnras/stx2687},
       eprint = {1610.01164}
}

@ARTICLE{Dijkstra2016a,
       author = {{Dijkstra}, Mark and {Sethi}, Shiv and {Loeb}, Abraham},
        title = "{3-cm Fine Structure Masers: A Unique Signature of Supermassive Black Hole Formation via Direct Collapse in the Early Universe}",    
      journal = {Astrophysical Journal},
         year = 2016,
       volume = {820},
        pages = {10},
          doi = {10.3847/0004-637X/820/1/10}
}

@ARTICLE{Dijkstra2016b,
       author = {{Dijkstra}, Mark and {Gronke}, Max and {Sobral}, David},
        title = "{Ly{\ensuremath{\alpha}} Signatures from Direct Collapse Black Holes}",
      journal = {Astrophysical Journal},
         year = 2016,
       volume = {823},
        pages = {74},
          doi = {10.3847/0004-637X/823/2/74}
}

@ARTICLE{Seon2020,
       author = {{Seon}, Kwang-il and {Kim}, Chang-Goo},
        title = "{Ly{\ensuremath{\alpha}} Radiative Transfer: Monte Carlo Simulation of the Wouthuysen-Field Effect}",
      journal = {Astrophysical Journal Supplement Series},
         year = 2020,
       volume = {250}, 
        pages = {9},
          doi = {10.3847/1538-4365/aba2d6}
}

@ARTICLE{Tang2026,
       author = {{Tang}, Mengtao and {Stark}, Daniel P. and {Mason}, Charlotte A. and {Chen}, Zuyi and {Katz}, Harley and {Gronke}, Max and {Furtak}, Lukas J. and {Chang}, Seok-Jun and {Matthee}, Jorryt and others},
        title = "{SPURS: Evidence for Clumpy Neutral Envelopes and Ionized IGM Surrounding Little Red Dots in Abell 2744 from Ultra-Deep Rest-UV Spectroscopy}",
      journal = {arXiv e-prints},
         year = 2026,
        pages = {arXiv:2604.03563},
          doi = {10.48550/arXiv.2604.03563},
       eprint = {2604.03563}
}

@ARTICLE{Naidu2026,
       author = {{Naidu}, Rohan P. and {Matthee}, Jorryt and {de Graaff}, Anna and {Torralba}, Alberto and {Ashall}, Chris and {Katz}, Harley and {Chisholm}, John and {Brammer}, Gabriel and {Dessart}, Luc and {Eilers}, Anna-Christina and others},
        title = "{Little Red Dots as Intermediate Mass, Super-Eddington Engines: Insights from Type IIn Supernovae and the 1837-1856 Great Eruption of {\ensuremath{\eta}} Carinae}",
      journal = {arXiv e-prints},
         year = 2026,
        pages = {arXiv:2606.30711},
          doi = {10.48550/arXiv.2606.30711},
       eprint = {2606.30711}
}

@ARTICLE{Chisholm2026,
       author = {{Chisholm}, John and {Berg}, Danielle A. and {Boylan-Kolchin}, Michael and {de Graaff}, Anna and {Furtak}, Lukas J. and {Kokorev}, Vasily and {Matthee}, Jorryt and {Mu{\~n}oz}, Julian B. and {Naidu}, Rohan P. and {Sander}, Andreas A.~C.},
        title = "{Little Red Dots as Globular Clusters in Formation}",
      journal = {arXiv e-prints},
         year = 2026,
        pages = {arXiv:2602.15935},
          doi = {10.48550/arXiv.2602.15935},
       eprint = {2602.15935}
}

@ARTICLE{Chon2026a,                                                                                                                           
       author = {{Chon}, Sunmyon and {Hirano}, Shingo and {Ishiyama}, Tomoaki and {Chang}, Seok-Jun and {Springel}, Volker},
        title = "{Overmassive black holes and little red dots naturally form in simulations}",
      journal = {Nature},
       volume = {657},
         year = 2026,
        pages = {621},
          doi = {10.1038/s41586-026-10985-8},
       eprint = {2601.04955}
}

@ARTICLE{Chon2026b,
       author = {{Chon}, Sunmyon and {Hirano}, Shingo and {Chen}, Ke-Jung and {Springel}, Volker},
        title = "{Formation of Heavy Seed Black Holes and Little Red Dots-like Compact Clusters in Metal-enriched Star-forming Regions}",
      journal = {arXiv e-prints},
         year = 2026,
        pages = {arXiv:2609.00437},
          doi = {10.48550/arXiv.2609.00437},
       eprint = {2609.00437}
}

@BOOK{Draine2011,
       author = {{Draine}, Bruce T.},
        title = "{Physics of the Interstellar and Intergalactic Medium}",
    publisher = {Princeton University Press},
         year = 2011,
      address = {Princeton, NJ},
         isbn = {9780691122144},
       series = {Princeton Series in Astrophysics},
}

@BOOK{OsterbrockFerland2006,
       author = {{Osterbrock}, Donald E. and {Ferland}, Gary J.},
        title = "{Astrophysics of Gaseous Nebulae and Active Galactic Nuclei}",
      edition = {2nd},
    publisher = {University Science Books},
         year = 2006,
      address = {Sausalito, CA},
}

@ARTICLE{Neufeld1990,
       author = {{Neufeld}, David A.},
        title = "{The Transfer of Resonance-Line Radiation in Static Astrophysical Media}",
      journal = {Astrophysical Journal},
         year = 1990,
        month = feb,
       volume = {350},
        pages = {216},
          doi = {10.1086/168375},
}

@ARTICLE{Dijkstra2014,
       author = {{Dijkstra}, Mark},
        title = "{Ly{\ensuremath{\alpha}} Emitting Galaxies as a Probe of Reionisation}",
      journal = {Publications of the Astronomical Society of Australia},
         year = 2014,
       volume = {31},
        pages = {e040},
          doi = {10.1017/pasa.2014.33},
}

@ARTICLE{Liu2025,
       author = {{Liu}, H. and {Jiang}, Y.-F. and {Quataert}, E. and {Greene}, J.~E. and {Ma}, Y.},
        title = "{The Balmer Break and Optical Continuum of Little Red Dots from Super-Eddington Accretion}",
      journal = {ApJ},
       volume = {994},
         year = 2025,
        pages = {113},
          doi = {10.3847/1538-4357/ae0c19}
}

@ARTICLE{Wang2024,
       author = {{Wang}, Bingjie and {Leja}, Joel and {de Graaff}, Anna and
{Brammer}, Gabriel B. and {Weibel}, Andrea and {van Dokkum}, Pieter and
{Baggen}, Josephine F. W. and {Suess}, Katherine A. and {Greene}, Jenny E. and
{Bezanson}, Rachel and {Cleri}, Nikko J. and {Hirschmann}, Michaela and
{Labbe}, Ivo and {Matthee}, Jorryt and {McConachie}, Ian and {Naidu}, Rohan P.
and {Nelson}, Erica and {Oesch}, Pascal A. and {Setton}, David J. and
{Williams}, Christina C.},
        title = "{RUBIES: Evolved Stellar Populations with Extended Formation
Histories at z $\sim$ 7-8 in Candidate Massive Galaxies Identified with
JWST/NIRSpec}",
      journal = {\apjl},
         year = 2024,
       volume = {969},
       number = {2},
        pages = {L13},
          doi = {10.3847/2041-8213/ad55f7},
}

@ARTICLE{Torralba2026,
       author = {{Torralba}, A. and {Matthee}, J. and {Pezzulli}, G. and {Naidu}, R.~P. and {Ishikawa}, Y. and {Brammer}, G.~B. and {Chisholm}, J. and {de Graaff}, A. and {D'Eugenio}, F. and {Di Cesare}, C. and {Eilers}, A.-C. and {Greene}, J.~E. and {Gronke}, M. and {Iani}, E. and {Kokorev}, V. and {Kotiwale}, G. and {Kramarenko}, I. and {Ma}, Y. and {Mascia}, S. and {Navarrete}, B. and {Nelson}, E. and {Oesch}, P. and {Simcoe}, R.~A. and {Wuyts}, S.},
        title = "{The warm outer layer of a little red dot as the source of [Fe II] and collisional Balmer lines with scattering wings}",
      journal = {A\&A},
       volume = {707},
         year = 2026,
        pages = {A75},
          doi = {10.1051/0004-6361/202557537}
}

@ARTICLE{Kokorev2026GLIMPSE,
       author = {{Kokorev}, V. and {Chisholm}, J. and {Naidu}, R.~P. and {Fujimoto}, S. and {Atek}, H. and {Brammer}, G. and {Finkelstein}, S.~L. and {Akins}, H.~B. and {Berg}, D. and {Furtak}, L.~J. and {Fei}, Q. and {Hsiao}, T.~Y.-Y. and {Labb{\'e}}, I. and {Matthee}, J. and {Mu{\~n}oz}, J.~B. and {Oesch}, P.~A. and {Pan}, R. and {Rinaldi}, P. and {Saldana-Lopez}, A. and {Schaerer}, D. and {Volonteri}, M. and {Zitrin}, A.},
        title = "{The Deepest GLIMPSE of a Dense Gas Cocoon Enshrouding a Little Red Dot}",
      journal = {ApJ},
       volume = {1004},
         year = 2026,
        pages = {153},
          doi = {10.3847/1538-4357/ae4ed7}
}

@ARTICLE{Kocevski23,
       author = {{Kocevski}, Dale D. and {Onoue}, Masafusa and {Inayoshi}, Kohei and {Trump}, Jonathan R. and {Arrabal Haro}, Pablo and {Grazian}, Andrea and {Dickinson}, Mark and {Finkelstein}, Steven L. and {Kartaltepe}, Jeyhan S. and {Hirschmann}, Michaela and others},
        title = "{Hidden Little Monsters: Spectroscopic Identification of Low-mass, Broad-line AGNs at $z > 5$ with CEERS}",
      journal = {Astrophysical Journal Letters},
         year = 2023,
       volume = {954},
        pages = {L4},
          doi = {10.3847/2041-8213/ace5a0}
}

@ARTICLE{Matthee2024,
       author = {{Matthee}, Jorryt and {Naidu}, Rohan P. and {Brammer}, Gabriel and {Chisholm}, John and {Eilers}, Anna-Christina and {Goulding}, Andy and {Greene}, Jenny and {Kashino}, Daichi and {Labbe}, Ivo and {Lilly}, Simon J. and others},
        title = "{Little Red Dots: An Abundant Population of Faint Active Galactic Nuclei at z {\ensuremath{\sim}} 5 Revealed by the EIGER and FRESCO JWST Surveys}",
      journal = {Astrophysical Journal},
         year = 2024,
       volume = {963},
        pages = {129},
          doi = {10.3847/1538-4357/ad2345}
}

@ARTICLE{Greene24,
       author = {{Greene}, Jenny E. and {Labbe}, Ivo and {Goulding}, Andy D. and {Furtak}, Lukas J. and {Chemerynska}, Iryna and {Kokorev}, Vasily and {Dayal}, Pratika and {Volonteri}, Marta and {Williams}, Christina C. and {Wang}, Bingjie and others},
        title = "{UNCOVER Spectroscopy Confirms the Surprising Ubiquity of Active Galactic Nuclei in Red Sources at $z > 5$}",
      journal = {Astrophysical Journal},
         year = 2024,
       volume = {964},
        pages = {39},
          doi = {10.3847/1538-4357/ad1e5f}
}

@ARTICLE{Juodzbalis2024a,
       author = {{Juod{\v{z}}balis}, Ignas and {Ji}, Xihan and {Maiolino}, Roberto and {D'Eugenio}, Francesco and {Scholtz}, Jan and {Risaliti}, Guido and {Fabian}, Andrew C. and {Mazzolari}, Giovanni and {Gilli}, Roberto and {Prandoni}, Isabella and others},
        title = "{JADES - the Rosetta stone of JWST-discovered AGN: deciphering the intriguing nature of early AGN}",
      journal = {Monthly Notices of the Royal Astronomical Society},
         year = 2024,
       volume = {535},
        pages = {853-873},
          doi = {10.1093/mnras/stae2367}
}

@ARTICLE{Kokorev2024LRD,
       author = {{Kokorev}, Vasily and others},
        title = "{A census of photometrically selected little red dots at $4 < z < 9$ in JWST blank fields}",
      journal = {Astrophysical Journal},
         year = 2024,
       volume = {968},
        pages = {38}
}

@ARTICLE{Setton24,
       author = {{Setton}, David J. and {Greene}, Jenny E. and {de Graaff}, Anna and {Ma}, Yilun and {Leja}, Joel and {Matthee}, Jorryt and {Bezanson}, Rachel and {Boogaard}, Leindert A. and {Cleri}, Nikko J. and {Katz}, Harley and others},
        title = "{Little Red Dots at an Inflection Point: Ubiquitous ``V-Shaped'' Turnover Consistently Occurs at the Balmer Limit}",
      journal = {arXiv e-prints},
         year = 2024,
        pages = {arXiv:2411.03424},
          doi = {10.48550/arXiv.2411.03424},
       eprint = {2411.03424}
}

@ARTICLE{Labbe24,
       author = {{Labbe}, Ivo and {Greene}, Jenny E. and {Matthee}, Jorryt and {Treiber}, Helena and {Kokorev}, Vasily and {Miller}, Tim B. and {Kramarenko}, Ivan and {Setton}, David J. and {Ma}, Yilun and {Goulding}, Andy D. and others},
        title = "{An unambiguous AGN and a Balmer break in an Ultraluminous Little Red Dot at z=4.47 from Ultradeep UNCOVER and All the Little Things Spectroscopy}",
      journal = {arXiv e-prints},
         year = 2024,
        pages = {arXiv:2412.04557},
          doi = {10.48550/arXiv.2412.04557},
       eprint = {2412.04557}
}

@ARTICLE{Inayoshi2025,
       author = {{Inayoshi}, Kohei and {Maiolino}, Roberto},
        title = "{Extremely Dense Gas around Little Red Dots and High-redshift Active Galactic Nuclei: A Nonstellar Origin of the Balmer Break and Absorption Features}",
      journal = {Astrophysical Journal Letters},
         year = 2025,
       volume = {980},
        pages = {L27},
          doi = {10.3847/2041-8213/adaebd}
}

@ARTICLE{Inayoshi2025b,
       author = {{Inayoshi}, Kohei},
        title = "{Little Red Dots as the Very First Activity of Black Hole Growth}",
      journal = {Astrophysical Journal Letters},
         year = 2025,
       volume = {988},
        pages = {L22},
          doi = {10.3847/2041-8213/adea66}
}

@ARTICLE{Naidu2025,
       author = {{Naidu}, Rohan P. and {Matthee}, Jorryt and {Katz}, Harley and {de Graaff}, Anna and {Oesch}, Pascal and {Smith}, Aaron and {Greene}, Jenny E. and {Brammer}, Gabriel and {Weibel}, Andrea and {Hviding}, Raphael E. and others},
        title = "{A ``Black Hole Star'' Reveals the Remarkable Gas-Enshrouded Hearts of the Little Red Dots}",
      journal = {arXiv e-prints},
         year = 2025,
        pages = {arXiv:2503.16596},
          doi = {10.48550/arXiv.2503.16596},
       eprint = {2503.16596}
}

@ARTICLE{Rusakov2025,
       author = {{Rusakov}, V. and {Watson}, D. and {Nikopoulos}, G.~P. and {Brammer}, G. and {Gottumukkala}, R. and {Harvey}, T. and {Heintz}, K.~E. and {Nielsen}, R.~D. and {Sim}, S.~A. and {Sneppen}, A. and others},
        title = "{JWST's little red dots: an emerging population of young, low-mass AGN cocooned in dense ionized gas}",
      journal = {arXiv e-prints},
         year = 2025,
        pages = {arXiv:2503.16595},
          doi = {10.48550/arXiv.2503.16595},
       eprint = {2503.16595}
}

@ARTICLE{Sneppen2026,
       author = {{Sneppen}, A. and others},
        title = "{Inside the cocoon: a comprehensive explanation of the spectra of Little Red Dots}",
      journal = {arXiv e-prints},
         year = 2026,
        pages = {arXiv:2601.18864},
       eprint = {2601.18864}
}

@ARTICLE{Chang2024MgII,
       author = {{Chang}, S.-J. and {Gronke}, M.},
        title = "{Probing cold gas with Mg\,\textsc{ii} and Ly\,$\alpha$ radiative transfer}",
      journal = {Monthly Notices of the Royal Astronomical Society},
         year = 2024,
       volume = {532},
        pages = {3526--3555},
          doi = {10.1093/mnras/stae1664}
}

@ARTICLE{ChangScattering2026,
       author = {{Chang}, S.-J. and {Gronke}, M. and {Matthee}, J. and {Mason}, C.},
        title = "{Impact of resonance, Raman, and Thomson scattering on hydrogen line formation in Little Red Dots}",
      journal = {Monthly Notices of the Royal Astronomical Society},
         year = 2026,
       volume = {545},
        pages = {1-21}
}

@ARTICLE{Karzas1961,
       author = {{Karzas}, W.~J. and {Latter}, R.},
        title = "{Electron Radiative Transitions in a Coulomb Field.}",
      journal = {Astrophysical Journal Supplement Series},
         year = 1961,
       volume = {6},
        pages = {167},
          doi = {10.1086/190063}
}

@ARTICLE{Murray1995,
       author = {{Murray}, N. and {Chiang}, J. and {Grossman}, S.~A. and {Voit}, G.~M.},
        title = "{Accretion Disk Winds from Active Galactic Nuclei}",
      journal = {\apj},
         year = 1995,
        month = oct,
       volume = {451},
        pages = {498},
          doi = {10.1086/176238},
       adsurl = {https://ui.adsabs.harvard.edu/abs/1995ApJ...451..498M}
}

@ARTICLE{Elvis2000,
       author = {{Elvis}, Martin},
        title = "{A Structure for Quasars}",
      journal = {\apj},
         year = 2000,
        month = dec,
       volume = {545},
       number = {1},
        pages = {63-76},
          doi = {10.1086/317778},
archivePrefix = {arXiv},
       eprint = {astro-ph/0008064},
 primaryClass = {astro-ph},
       adsurl = {https://ui.adsabs.harvard.edu/abs/2000ApJ...545...63E}
}

@ARTICLE{Ohsuga2005,
       author = {{Ohsuga}, Ken and {Mori}, Masao and {Nakamoto}, Taishi and {Mineshige}, Shin},
        title = "{Supercritical Accretion Flows around Black Holes: Two-dimensional, Radiation Pressure-dominated Disks with Photon Trapping}",
      journal = {\apj},
         year = 2005,
        month = jul,
       volume = {628},
       number = {1},
        pages = {368-381},
          doi = {10.1086/430728},
archivePrefix = {arXiv},
       eprint = {astro-ph/0504168},
 primaryClass = {astro-ph},
       adsurl = {https://ui.adsabs.harvard.edu/abs/2005ApJ...628..368O}
}

@ARTICLE{Jiang2014,
       author = {{Jiang}, Yan-Fei and {Stone}, James M. and {Davis}, Shane W.},
        title = "{A Global Three-dimensional Radiation Magneto-hydrodynamic Simulation of Super-Eddington Accretion Disks}",
      journal = {\apj},
         year = 2014,
        month = dec,
       volume = {796},
       number = {2},
          eid = {106},
        pages = {106},
          doi = {10.1088/0004-637X/796/2/106},
archivePrefix = {arXiv},
       eprint = {1410.0678},
 primaryClass = {astro-ph.HE},
       adsurl = {https://ui.adsabs.harvard.edu/abs/2014ApJ...796..106J}
}

@ARTICLE{Matthee2026,
       author = {{Matthee}, Jorryt and {Torralba}, Alberto and {Pezzulli}, Gabriele and {Naidu}, Rohan P. and {Chisholm}, John and {Mascia}, Sara and {Greene}, Jenny E. and {Ishikawa}, Yuzo and {Gronke}, Max and {Wuyts}, Stijn and {Bordoloi}, Rongmon and {Brammer}, Gabriel and {Chang}, Seok-Jun and {Eilers}, Anna-Christina and {de Graaff}, Anna and {Hviding}, Raphael E. and {Iani}, Edoardo and {Illingworth}, Garth and {Kashino}, Daichi and {Labbe}, Ivo and {Ma}, Yilun and {Maseda}, Michael V. and {Meyer}, Romain and {Nelson}, Erica and {Oesch}, Pascal and {Xiao}, Mengyuan},
        title = "{The Engine and its Flows: Little Red Dot spectra are shaped by the column densities of their gas envelopes}",
      journal = {arXiv e-prints},
         year = 2026,
        month = mar,
          eid = {arXiv:2603.17667},
        pages = {arXiv:2603.17667},
          doi = {10.48550/arXiv.2603.17667},
archivePrefix = {arXiv},
       eprint = {2603.17667},
 primaryClass = {astro-ph.GA},
       adsurl = {https://ui.adsabs.harvard.edu/abs/2026arXiv260317667M}
}

@article{ForemanMackey2013,
  author  = {{Foreman-Mackey}, D. and {Hogg}, D.~W. and {Lang}, D. and {Goodman}, J.},
  title   = {{emcee}: The {MCMC} Hammer},
  journal = {PASP},
  year    = {2013},
  volume  = {125},
  pages   = {306--312},
  doi     = {10.1086/670067},
  eprint  = {1202.3665},
  primaryClass = {astro-ph.IM}
}

@article{Cameron2023,
  author  = {{Cameron}, A.~J. and {Saxena}, A. and {Bunker}, A.~J. and
             {D'Eugenio}, F. and {Carniani}, S. and {Maiolino}, R. and
             {Curtis-Lake}, E. and {Ferruit}, P. and {Jakobsen}, P. and
             {Arribas}, S. and {Charlot}, S. and {Chevallard}, J. and
             {Curti}, M. and {Looser}, T.~J. and {Maseda}, M.~V. and
             {Rawle}, T. and {Smit}, R. and {{{\"U}bler}}, H. and
             {Willott}, C. and {Witstok}, J. and {Eisenstein}, D.~J. and
             {Hainline}, K. and {Rieke}, M. and {Robertson}, B.~E. and
             {Stark}, D.~P. and {Tacchella}, S. and {Willmer}, C.~N.~A.},
  title   = {{JADES}: Probing interstellar medium conditions at $z\sim5.5$--9.5
             with ultra-deep {JWST/NIRSpec} spectroscopy},
  journal = {A\&A},
  year    = {2023},
  volume  = {677},
  pages   = {A115},
  doi     = {10.1051/0004-6361/202346107},
  eprint  = {2302.04298},
  primaryClass = {astro-ph.GA}
}

@ARTICLE{Laor2006,
       author = {{Laor}, Ari},
        title = "{Evidence for Line Broadening by Electron Scattering in the Broad-Line Region of NGC 4395}",
      journal = {\apj},
         year = 2006,
       volume = {643},
        pages = {112--119}
}

@ARTICLE{SJC2023,
       author = {{Chang}, Seok-Jun and {Yang}, Yujin and {Seon}, Kwang-Il and {Zabludoff}, Ann and {Lee}, Hee-Won},
        title = "{Radiative Transfer in Ly$\alpha$ Nebulae. I. Modeling a Continuous or Clumpy Spherical Halo with a Central Source}",
      journal = {Astrophysical Journal},
         year = 2023,
       volume = {945},
        pages = {100},
          doi = {10.3847/1538-4357/acac98}
}

@ARTICLE{Neufeld1991,
       author = {{Neufeld}, David A.},
        title = "{The transfer of resonance-line radiation in clumpy media}",
      journal = {Astrophysical Journal Letters},
         year = 1991,
       volume = {370},
        pages = {L85},
          doi = {10.1086/185983}
}

@ARTICLE{Gronke2016,
       author = {{Gronke}, Max and {Dijkstra}, Mark and {McCourt}, Michael and {Oh}, S. Peng},
        title = "{From Mirrors to Windows: Lyman-alpha Scattering around Galaxies in the Presence of Multiphase Gas}",
      journal = {Astrophysical Journal Letters},
         year = 2016,
       volume = {833},
        pages = {L26},
          doi = {10.3847/2041-8213/833/2/L26}
}

@ARTICLE{Gronke2017,
       author = {{Gronke}, Max and {Dijkstra}, Mark and {McCourt}, Michael and {Oh}, S. Peng},
        title = "{Resonance line transfer in multiphase clumpy media}",
      journal = {Astronomy and Astrophysics},
         year = 2017,
       volume = {607},
        pages = {A71},
          doi = {10.1051/0004-6361/201731055}
}

\section*{Methods}\label{sec:methods}

\subsection*{Sample selection and spectroscopy}\label{subsec:meth_sample}

Our sample of 11 LRDs is drawn from the broad H$\alpha$-selected sample of \cite{Matthee2026}, which compiled high-quality JWST NIRSpec spectra at $z \approx 3.5$--7 from multiple JWST survey programs. Both NIRSpec PRISM/CLEAR and higher-resolution grating (G235M or G395H) spectra are available for all objects; we use the PRISM spectra for the SED continuum fitting and the grating spectra for the Balmer-line profile modeling. We select the 11 objects with sufficient signal-to-noise in both \ha and \hb for joint line-profile fitting; the sample is listed in Extended Data Table~\ref{tab:sample}. Full details of the sample selection and data reduction are described in \cite{Matthee2026}.

\subsection*{Balmer-line radiative-transfer modelling}\label{subsec:meth_rt}

We model the H$\alpha$ and H$\beta$ line profiles using \textsc{rt-scat}, a Monte Carlo radiative transfer code for line formation by scattering processes previously applied to Ly$\alpha$\cite{SJC2023} and Mg\,\textsc{ii} emission\cite{Chang2024MgII}; the RT algorithm and clumpy-medium treatment are described in detail therein.
The Balmer-line implementation is described in \cite{ChangScattering2026}, which is our previous work on Balmer-line radiative transfer for resonance and Thomson scattering\footnote{S.-J. Chang, the lead author of this work, is a developer of \textsc{rt-scat} for resonance scattering and its application to Balmer lines}. The scattering geometry consists of a central isotropic source surrounded by two concentric spherical regions (Fig.~15 of \cite{ChangScattering2026}; illustrated for both models in Extended Data Fig.~\ref{fig:ext_geometry}): an inner, fully ionised H\,\textsc{ii} region ($r < 0.01\,R$) responsible for Thomson scattering, characterised by electron temperature $T_{\rm e} = 10^4$\,K and Thomson optical depth $\tau_{\rm Th}$; and an outer neutral H\,\textsc{i} region ($0.01\,R < r < R$) with a significantly populated $n=2$ level, responsible for Balmer-line resonance scattering. Photons emitted by the central source first traverse the Thomson-scattering H\,\textsc{ii} layer, which broadens the line wings symmetrically, before entering the outer H\,\textsc{i} layer where resonance scattering shapes the line core and absorption features. We consider two configurations for the outer H\,\textsc{i} sphere: a \textit{homogeneous} sphere with uniform $n=2$ \hi column density $N_{\rm HI,\mathit{n}=2}$, and a \textit{clumpy} sphere characterised by the same total column density $N_{\rm HI,\mathit{n}=2}$ distributed in spherical optically-thick clumps of fixed radius $r_{\rm cl} = 10^{-3}\,R$, embedded in an optically thin inter-clump medium; the individual clump column density is $N_{\rm HI,cl} = \tfrac{3}{4}\,N_{\rm HI,\mathit{n}=2}/f_c$ (the factor $3/4$ accounts for the mean chord length through a sphere); the covering factor $f_c$ is the mean number of clumps along a line of sight from the central source to the outer edge of the sphere\cite{Chang2024MgII}. 

The H$\alpha$ and H$\beta$ resonance cross-sections at line center differ by a factor of $\approx 5.7$ (the ratio of oscillator strengths times wavelength; \cite{ChangScattering2026} equations~3 and~4), so that H$\alpha$ photons experience substantially higher opacity than H$\beta$ at the same $N_{\rm HI,\mathit{n}=2}$. In the clumpy case, H$\alpha$ photons scatter predominantly off the near surface of each clump and escape without penetrating to high column density (\textit{surface scattering}), whereas H$\beta$ photons penetrate more deeply and are subject to a branching decay at the $4p$ level ($\rm{BR}_{n=3} \approx 0.26$; \cite{ChangScattering2026} Table~3 and Fig.~4) that removes them from the H$\beta$ line before re-emission. This mechanism allows the clumpy model to simultaneously reproduce both lines with a single $N_{\rm HI,\mathit{n}=2}$, including cases where H$\beta$ absorption is deeper than H$\alpha$. The asymmetry is atomic in origin: the $3p \to 1s$ (Ly$\beta$) decay is immediately reabsorbed by ground-state hydrogen, so a scattered H$\alpha$ photon is ultimately re-emitted as H$\alpha$, whereas the $4p$ level also decays to $3s$ and $3d$, which cascade through $2p$ and emit Pa$\alpha$ and H$\alpha$ instead of H$\beta$. With ${\rm BR}_3 \approx 0.26$ per scattering, $1-(1-{\rm BR}_3)^4 \approx 70\%$ of H$\beta$ photons are converted after four scatterings.

The free parameters of the homogeneous model are $N_{\rm HI,\mathit{n}=2}$, the radial velocity $v_{\rm exp}$, the intrinsic emission-line width $\sigma_{\rm emit}$, the random velocity dispersion $\sigma_{\rm ran}$, and the Thomson optical depth $\tau_{\rm Th}$. The clumpy model retains $N_{\rm HI,\mathit{n}=2}$ (total column density) and adds the covering factor $f_c$; the per-clump column density $N_{\rm HI,cl} = \tfrac{3}{4}\,N_{\rm HI,\mathit{n}=2}/f_c$ is a derived quantity. 
In the clumpy model, $\sigma_{\rm ran}$ is the peculiar velocity dispersion of individual clumps about the constant outflow at $v_{\rm exp}$, rather than a Doppler width of the bulk medium as in the homogeneous case; the random motion within clumps is the thermal motion, as their temperature is at $T=10^4\, \rm K$.
Parameter ranges are given in Extended Data Table~\ref{tab:rt_params}. The intrinsic emission width is restricted to $\sigma_{\rm emit} \in [100, 300]$\,\kms: larger values would attribute the broad line wings to intrinsic emission rather than Thomson scattering, removing the physical motivation for the two-component geometry of \cite{ChangScattering2026}. 
The total grid comprises 40,130 homogeneous and 181,212 clumpy model spectra.
The \NHItwo lower bound of $10^{12}$\,cm$^{-2}$ represents optically thin gas, where $\tau_{0,{\rm H}\alpha}\approx0.3$ and $\tau_{0,{\rm H}\beta}\approx0.06$ (equations~3 and~4 of \cite{ChangScattering2026}). Thus, the gas is effectively transparent in both lines, and the fit is insensitive to $N_{\rm HI,\mathit{n}=2}$ below this value.

The continuum is first removed from each spectrum using the continuum model of \cite{Matthee2026}, derived from a spectral decomposition of the full JWST NIRSpec grating spectrum. The narrow host-galaxy emission in H$\alpha$ and H$\beta$ is then subtracted. For each object, the [O\,\textsc{iii}]\,$\lambda5007$ profile is used as a host template for both Balmer lines: $F_{\rm host,H\beta}(v) = F_{\rm [OIII]}(v)/R_{\rm H\beta}$ and $F_{\rm host,H\alpha}(v) = F_{\rm [OIII]}(v)/R_{\rm H\alpha}$, where $R_{\rm H\beta} \equiv F_{\rm [OIII]}/F_{\rm H\beta,host}$ and $R_{\rm H\alpha} \equiv F_{\rm [OIII]}/F_{\rm H\alpha,host}$ are the host [O\,\textsc{iii}]-to-Balmer line ratios. Both are free parameters. For $R_{\rm H\beta}$ we adopt a Gaussian prior on $\log_{10} R_{\rm H\beta}$ centered on the median [O\,\textsc{iii}]/H$\beta$ ratio of high-z galaxies at the redshift of each object \cite{Cooper2026}, with a width of 0.15\,dex \cite{Cameron2023}, truncated so that the subtracted host does not drive the observed H$\beta$ and H$\alpha$ line cores below zero beyond their observational uncertainties. The host Balmer decrement, $R_{\rm H\beta}/R_{\rm H\alpha}$, is allowed to vary between 2.6 and 3.3 (Case~B with mild departures). A third free parameter, $\Delta v$, is a single velocity offset shared by both lines between the [O\,\textsc{iii}]-derived host/narrow-line-region zero point and the true center of the LRD's own broad-line/clumpy-gas structure; it has a Gaussian prior centered at zero with width set in quadrature from that object's own [O\,\textsc{iii}] line width and the instrumental resolution.

We infer posterior distributions for all free parameters using Markov chain Monte Carlo (MCMC) sampling via the affine-invariant ensemble sampler \textsc{emcee}\cite{ForemanMackey2013}. At each proposed parameter vector, the model line profiles are evaluated at continuous parameter values via a principal-component-analysis emulator trained on the precomputed \textsc{rt-scat} grid, allowing the sampler to explore the full parameter space between grid points. Each $\chi^2$ is evaluated over the velocity windows $|v| \leq 3000$\,\kms (H$\alpha$) and $|v| \leq 2000$\,\kms (H$\beta$) using an optimal linear normalisation $A = (\sum_i O_i M_i \sigma_i^{-2}) / (\sum_i M_i^2 \sigma_i^{-2})$ sensitive only to line profile shape. The per-pixel uncertainty is $\sigma_i^2 = \sigma_{{\rm obs},i}^2 + \sigma_{{\rm jig},i}^2$, where $\sigma_{\rm obs}$ is the pipeline error and $\sigma_{{\rm jig},i} = \sigma_{\rm MAD}\sqrt{F_{\rm ref}\,|f_i|}$ is a conservative allowance for residual spectral structure not captured by it. Here $\sigma_{\rm MAD}$ is the scatter of each line's wings ($1500<|v|<5000$\,\kms) about a symmetric exponential profile, commonly used to describe the broad wings of LRD Balmer lines \cite{Laor2006,Rusakov2025}, and $F_{\rm ref}$ is the median wing flux at $3000<|v|<5000$\,\kms. Model spectra are convolved to the instrumental resolution prior to comparison ($\mathcal{R} = 2700$ for G235H/G395H, $\mathcal{R} = 1000$ for G235M/G395M).

We fit H$\alpha$ and H$\beta$ in two sequential stages, taking H$\alpha$ as the baseline because it has the highest signal-to-noise ratio of the Balmer lines in our sample and its profile most directly encodes the scattering geometry.  
\begin{itemize}
    \item 
\textit{Stage 1:} we fit the H$\alpha$ profile alone with flat priors on all radiative transfer parameters (Extended Data Table~\ref{tab:rt_params}). Because H$\beta$ photons that decay through $n=3$ are partly re-emitted as H$\alpha$, the model H$\alpha$ profile includes this converted component, weighted by the intrinsic H$\alpha$/H$\beta$ flux ratio of the source, $F_{\rm H\alpha}/F_{\rm H\beta}$, a free parameter with a Gaussian prior centered on the observed $F_{\rm H\alpha}/F_{\rm H\beta}$ of each object. We use a kernel density estimate of the resulting posterior on $\log N_{\rm HI,\mathit{n}=2}$ as the prior for the next stage.

\item
\textit{Stage 2:} we fit the H$\beta$ profile alone with this prior imposed on $\log N_{\rm HI,\mathit{n}=2}$. All other radiative transfer parameters retain flat priors and are refit independently of Stage~1. This tests whether the \NHItwo from the H$\alpha$ fit in Stage~1 is also consistent with the observed H$\beta$ profile.
\end{itemize}

For each stage, we run 128 walkers for 50,000 steps in Stage~1 and 64 walkers for 20,000 steps in Stage~2 per object per model, initialized at random positions within the prior ranges; the first 5,000 steps are discarded as burn-in, and marginalized 68\% credible intervals are derived from the remaining samples. Full spectral fits for all objects are shown in Extended Data Fig.~\ref{fig:ext_fit1} and \ref{fig:ext_fit2}. Marginalized $\log\,N_{\rm HI,\mathit{n}=2}$ posteriors are shown in Extended Data Fig.~\ref{fig:ext_nhi_post_clumpy} and \ref{fig:ext_nhi_post_smooth}. Fitted parameters for all 11 objects are listed in Extended Data Tables~\ref{tab:ext_smooth} and~\ref{tab:ext_clumpy}. 

Three effects limit the constraints for individual objects. Host subtraction can affect the line cores, which is why the host Balmer contributions are treated as free parameters with observational priors rather than fixed values. At $\mathcal{R}\sim1000$, the H$\beta$ core is unresolved, which broadens the \NHItwo posteriors. In the clumpy model, H$\alpha$ is only weakly sensitive to \NHItwo because photons scatter off the clump surfaces, so \NHItwo is partly degenerate with $f_c$ and its credible intervals are typically wider than in the homogeneous model ($\sim$2 versus $\sim$1\,dex; Fig.~\ref{fig:nhi_comparison} and Extended Data Tables~\ref{tab:ext_smooth} and \ref{tab:ext_clumpy}), particularly when no clear absorption feature is detected. H$\beta$ remains sensitive to \NHItwo; the Stage~2 fit therefore combines the constraints of both lines and, for most objects, narrows the \NHItwo interval, which is why we use it in Fig.~\ref{fig:nhi_color}.

\subsection*{Balmer-break SED modelling}\label{subsec:meth_sed}

We model the rest-frame SED of each LRD with three spectral components. The first is a UV power-law continuum, $f_\lambda \propto \lambda^{\alpha_{\rm UV}}$. The second is an optical continuum component, modeled as a blackbody with effective temperature $T_{\rm BB}$. In practice, bound-free (Balmer and Paschen recombination continua) and free-free emission from the ionized gas can also contribute to the optical continuum alongside the blackbody component; $T_{\rm BB}$ should therefore be understood as an effective fitting parameter rather than a physical temperature. The third component is a Balmer break applied to the optical continuum blueward of the Balmer limit $\lambda_{\rm BE} = 3646$\,\AA, with bound-free optical depth scaling as $\tau_{\rm bc}(\lambda) = \tau_0\,(\lambda/\lambda_{\rm BE})^3$, where $\tau_0 = N_{\rm HI,\mathit{n}=2}\,\sigma_{\rm bc}$ and $\sigma_{\rm bc} = 1.5\times10^{-17}$\,cm$^2$\,\cite{Karzas1961}. The UV power-law component is left unattenuated, while only the optical continuum component is subject to this attenuation. This differential treatment is a modeling choice motivated by the fit itself: the optical continuum must arise from gas that shares the same $n=2$ \hi absorbing column responsible for the Balmer break and Balmer-line absorption, whereas the UV power-law component does not --- consistent with it originating from a source not co-located with the $n=2$ \hi clumps (e.g., unobscured host-galaxy light). Broad emission lines are excluded from the fit by adopting continuum windows that avoid all major rest-frame emission features (see below).

The fit proceeds in three sequential stages. 
\begin{itemize}
    \item 
\textit{Stage~1:} the UV power-law amplitude and slope $\alpha_{\rm UV}$ are determined from emission-line-free UV windows (rest-frame 1300--1420, 2000--2300, and 3400--3620\,\AA)\cite{Matthee2026}.

\item
\textit{Stage~2:} the Stage 1 UV SED is subtracted from the observed spectrum and the residual optical continuum is fitted with the blackbody component in line-free optical windows (3655--4000, 4230--4280, 5100--5800, 6000--6270, and 7200--8000\,\AA)\cite{Matthee2026}; each window receives equal weight regardless of pixel count to prevent the fit being dominated by wide wavelength regions.

\item
\textit{Stage~3:} with the UV and optical SEDs frozen from Stages~1 and~2, we attenuate the combined continuum with the Balmer-continuum optical depth ($\propto \nu^{-4}$) described above blueward of 3646\,\AA, plus a sum of Voigt-profile line optical depth from the higher Balmer series (upper level $n=3$--100, each weighted by its oscillator strength and total spontaneous decay rate) redward of 3646\,\AA, sharing a common Doppler width $v_{\rm ran}$. Both terms are evaluated over the window 3000--3800\,\AA, which brackets the Balmer limit on both sides and is sensitive to both the depth of the break and the level of the unattenuated optical continuum just redward of 3646\,\AA. We mask 3700--3760\,\AA, where [O\,\textsc{ii}]\,$\lambda3727$ blends with the high-$n$ Balmer lines at PRISM resolution. The two free parameters are $N_{\rm HI,\mathit{n}=2}$, which sets both optical depth terms, and $v_{\rm ran} \in [10,500]$\,\kms. Rather than a flat prior, $\log N_{\rm HI,\mathit{n}=2}$ is given the same prior constructed from the H$\alpha$ fit posterior, while $v_{\rm ran}$ has a flat prior. The flux uncertainties include a residual-scatter term of the same form as for the Balmer lines, measured from the residuals about the Stage~1/2 continuum. We sample the posterior with \textsc{emcee} (32 walkers, 12,000 steps, the first 2,000 discarded), with the Stage~1/2 continuum held fixed.
\end{itemize}

We quantify the relative prominence of the optical continuum component through the optical-excess parameter
\begin{equation}\label{eq:fexc}
f_{\rm exc} = \frac{\langle f_{\rm obs} \rangle_{\rm opt} - \langle f_{\rm UV} \rangle_{\rm opt}}{\langle f_{\rm UV} \rangle_{\rm opt}},
\end{equation}
where $\langle f_{\rm obs} \rangle_{\rm opt}$ and $\langle f_{\rm UV} \rangle_{\rm opt}$ are the mean observed flux density and mean UV power-law model flux density averaged over the Stage-2 optical windows, respectively. We claim a Balmer-break detection only when two criteria are satisfied simultaneously: (i)~$f_{\rm exc} \geq 1$, ensuring that a genuine red continuum component exists against which a break can be measured; and (ii)~$\Delta{\rm BIC} > 0$, computed with the pipeline flux uncertainties, meaning the Stage-3 model is preferred over the Stage-2 model without attenuation. Of the 11 LRDs in our sample, 9 satisfy both criteria. The remaining two --- GS~3073 ($f_{\rm exc} = 0.36$) and JADES~GN~38147 ($f_{\rm exc} = 0.63$) --- have UV-dominated SEDs in which the optical continuum is too faint relative to the inner-disc UV to anchor a reliable Balmer-break measurement. 
These two non-detections do not imply an absence of $n=2$ \hi gas for SEDs.

Because both the Balmer break and the higher Balmer series lines are included in the Stage 3 fit, we constrain \NHItwo and $v_{\rm ran}$ directly, rather than providing only a saturation-limited lower bound on \NHItwo as a continuum-only fit would. For a small number of objects, the fit remains prior-dominated (see flags in Extended Data Table~\ref{tab:sed_fits}). Best-fit spectra for all objects are shown in the third column of Extended Data Fig.~\ref{fig:ext_fit1} and \ref{fig:ext_fit2} (for homogeneous fits that cannot produce the break, see Fig.~\ref{fig:example_fit} caption); full posterior distributions are shown in Extended Data Fig.~\ref{fig:ext_nhi_post_clumpy} and \ref{fig:ext_nhi_post_smooth}, with numerical values listed in Extended Data Table~\ref{tab:sed_fits}. Full posterior distributions from both stages for GN~9771 and A2744~45924 are shown in Extended Data Figs.~\ref{fig:ext_corner_gn9771} and \ref{fig:ext_corner_a2744}.

\subsection*{Line--break consistency test}\label{subsec:meth_consistency}

For each of the 9 LRDs with a Balmer-break detection, we test whether the \NHItwo from the H$\alpha$ fit can also produce the Balmer break. A model fails if either (i)~the median H$\alpha$ \NHItwo lies more than a factor of 100 below the Balmer-break threshold, $N_{\rm HI,\mathit{n}=2} \approx 6\times10^{16}$\cmtwo ($\tau_{\rm BB}=1$), i.e. $\tau_{\rm BB} < 0.01$, for which the break depth would be below one percent; or (ii)~the 84th percentile of the SED fit with the H$\alpha$ prior lies below $10^{16.5}$\cmtwo ($\tau_{\rm BB}=0.5$), i.e. the fit cannot produce the break. 
Failures under (i) are robust to the posterior width for 3 of the 6 homogeneous objects that fail it (GN~9771, A2744~45924, and RUBIES~UDS~154183): the 84th percentile of the H$\alpha$ \NHItwo also lies more than two orders of magnitude below the threshold. Of the 9 detected objects, the clumpy model passes for all 9, and the homogeneous model passes for 3 (RUBIES~EGS~55604, RUBIES~EGS~49140, and GN~15498). For RUBIES~EGS~49140 and GN~15498, the homogeneous H$\alpha$ trough sits at the line center ($v_{\rm exp}\approx10$--$20$\,\kms), where its depth depends on the host subtraction, and the H$\alpha$ \NHItwo posteriors are broad (68\% intervals spanning 2.6 and 3.3\,dex). Results are summarized in Fig.~\ref{fig:nhi_comparison} and Extended Data Table~\ref{tab:sed_fits}.

At \NHItwo from the H$\alpha$ fit (Stage 2), the clumpy model reproduces the H$\beta$ absorption feature, whereas the homogeneous model predicts absorption that is too weak, for GN~9771, A2744~45924, and RUBIES~UDS~154183  (Fig.~\ref{fig:example_fit} and Extended Data Fig.~\ref{fig:ext_fit1}); the reduced $\chi^2$ of the clumpy fit is lower by factors of 1.4--3.2 (Extended Data Tables~\ref{tab:ext_smooth} and \ref{tab:ext_clumpy}). For the other objects, the two fits are comparable ($\chi^2$ ratios within 1.0--1.04).
For RUBIES~EGS~42046, however, the homogeneous H$\beta$ fit shows $\log\NHItwo\approx17.6$, nearly three orders of magnitude above its H$\alpha$ value, so the tension there is between the two lines rather than with the break (Fig.~\ref{fig:nhi_comparison}b); the clumpy fits agree within their uncertainties.

Marginalized posterior distributions for all probes (H$\alpha$, H$\beta$, and SED, each with and without the H$\alpha$-derived prior) are shown in Extended Data Fig.~\ref{fig:ext_nhi_post_clumpy} and \ref{fig:ext_nhi_post_smooth} for clumpy and homogeneous models, respectively.


\subsection*{Nebular continuum and H$\alpha$ intrinsic equivalent width}

Extended Data Fig.~\ref{fig:ext_neb_bb_ew} quantifies the equivalent width (EW) of \ha that a purely nebular optical continuum would imply, before any optically thick \ha radiative-transfer effects. This sets the baseline against which the observed LRD \ha EWs, and any proposed \ha suppression, must be judged.

We compute the hydrogen nebular continuum as the sum of free--free emission and the free--bound (radiative recombination) continua to $n=2$ and $n=3$,
\begin{equation}
j_\lambda^{\rm neb} = j_\lambda^{\rm ff} + j_\lambda^{\rm fb}(n\!=\!2) + j_\lambda^{\rm fb}(n\!=\!3),
\end{equation}
evaluated over rest-frame $1500$--$8000$\,\AA. Free-bound recombination to $n\geq4$ contributes $\lesssim3$ percent of the continuum in this range and is neglected. Recombination coefficients are from the tabulation of \cite{Draine2011}. 
Two photon $2s\rightarrow1s$ emission is omitted: at the densities invoked for LRD line-emitting gas in previous modeling ($n_e\gtrsim10^{5}$\,cm$^{-3}$, well above the $2s$ critical density), collisional $2s\leftrightarrow2p$ mixing redirects these decays into Ly$\alpha$; adding a two-photon term computed at $n_e=10^{9}$\,cm$^{-3}$ changes $\mathrm{EW}_{\mathrm{H}\alpha}$ by $<0.01$ percent at every temperature tested. The gas is taken to be photoionized and in Case~B recombination \cite{OsterbrockFerland2006}, with bound-level populations taken in the low-density, collisionless limit apart from the $2s\leftrightarrow2p$ mixing noted above, and \ha is treated as optically thin in this baseline emissivity calculation.

The intrinsic rest-frame EW is
\begin{equation}
\mathrm{EW}_{\mathrm{H}\alpha}^{\rm int}=\frac{L_{\mathrm{H}\alpha}}{L_{\lambda,\rm neb}(\lambda_{\mathrm{H}\alpha})},
\end{equation}
with $L_{\mathrm{H}\alpha}=n_e n_p\,\alpha_{\mathrm{H}\alpha}^{\rm eff}(T)\,h\nu_{\mathrm{H}\alpha}\,V$. Both the line and the continuum luminosity scale with the emission measure $\int n_e n_p\,\mathrm{d}V$, so the electron density and emitting volume cancel in the ratio and $\mathrm{EW}_{\mathrm{H}\alpha}^{\rm int}$ depends only on temperature, through the atomic coefficients. The adopted fiducial density ($n_e=n_p=10^{9}$\,cm$^{-3}$) is therefore immaterial to the result; it is motivated by previous LRD modeling rather than by the present Balmer radiative-transfer fits. We evaluate $\mathrm{EW}_{\mathrm{H}\alpha}^{\rm int}$ on a grid of $T=10^{3}$--$10^{5}$\,K.

Extended Data Fig.~\ref{fig:ext_neb_bb_ew} is built from this calculation. The left panel shows the nebular continuum at $T=5000$, $10^{4}$, and $2\times10^{4}$\,K, each normalized to unity at rest-frame 5500\,\AA, with \ha and \hb drawn on top as Gaussian emission lines (velocity dispersion $500$\,\kms) whose integrated fluxes are set self-consistently from $\mathrm{EW}_{\mathrm{H}\alpha}^{\rm int}$ at that temperature and the Case~B ratio $\mathrm{H}\alpha/\mathrm{H}\beta=2.86$. The right panel plots $\mathrm{EW}_{\mathrm{H}\alpha}^{\rm int}$ against $T$; the shaded band marks the representative observed LRD range, $\mathrm{EW}_{\mathrm{H}\alpha}\sim100$--$2000$\,\AA\ \cite{Matthee2026}.

Over the plotted grid $\mathrm{EW}_{\mathrm{H}\alpha}^{\rm int}$ is of order $10^{3}$--$10^{5}$\,\AA, and equals $\approx1.4\times10^{4}$, $9.1\times10^{3}$, and $5.7\times10^{3}$\,\AA\ at $T=5000$, $10^{4}$, and $2\times10^{4}$\,K --- roughly $10$--$100$ times larger than the observed LRD \ha EWs. A nebular-dominated optical continuum therefore overpredicts the \ha equivalent width, and reconciling the two requires either substantial \ha suppression or a separate, non-nebular optical-continuum component. Because both the \ha line and the nebular continuum scale with the emission measure, this equivalent width is set by atomic physics alone --- independent of density, geometry, and the number of emitting zones --- and it stays $\gtrsim10^{3}$\,\AA\ across the whole temperature range, so a multi-temperature mixture cannot evade it either. Any model that reproduces the observed, lower \ha equivalent widths from nebular gas --- including radiative-transfer treatments of dense LRD envelopes --- must therefore invoke one of the two routes above: additional suppression of \ha, or an optical continuum that is not purely recombination emission. Although the red color of the nebular continuum is set by fixed atomic thresholds (the Balmer edge at 3646\,\AA\ and the Paschen edge at 8204\,\AA), its slope and break strength still depend on temperature; the continuum shape is not temperature-independent.

The predicted EW also assumes that \ha is not collisionally suppressed. Collisional de-excitation of the $n=3$ level becomes significant only near its critical density $n_{\rm crit}=A/q$ \cite{Draine2011}, where $A$ is the Einstein A coefficient and $q$ the collisional de-excitation rate coefficient. Averaged over the $n=3$ sublevels the two radiative channels are $A_{32}\approx4.4\times10^{7}$\,s$^{-1}$ (\ha) and $A_{31}\approx5.6\times10^{7}$\,s$^{-1}$ (Ly$\beta$), so with $q\approx1.6\times10^{-7}$\,cm$^{3}$\,s$^{-1}$ at $T=10^{4}$\,K the critical density is $n_{\rm crit}\approx6\times10^{14}$\,cm$^{-3}$ if both photons escape. In gas optically thick to the Lyman series, the $3\rightarrow1$ decay does not remove the excitation: the Ly$\beta$ photon is reabsorbed locally, since the Lyman lines scatter off the ground state and are more optically thick still, and the atom is returned to $n=3$, so that a photon alternates in identity between Ly$\beta$ and \ha until it escapes as \ha\cite{Chang2018}. The radiative loss rate from $n=3$ is then set by $A_{32}$ alone and $n_{\rm crit}\approx3\times10^{14}$\,cm$^{-3}$, varying by less than a factor of two over $T=3000$--$2\times10^{4}$\,K. Including the collisional $3\rightarrow n>3$ channels, which depopulate $n=3$ without necessarily destroying the \ha photon, lowers this to $\sim4\times10^{13}$\,cm$^{-3}$. All of these lie well above the densities invoked for LRD line-emitting gas in previous studies, $n_{\rm H}\sim10^{9}$--$10^{11}$\,cm$^{-3}$\cite{Naidu2025,Inayoshi2025}, and above the $10^{6}$--$10^{10}$\,cm$^{-3}$ adopted where the optical continuum is attributed to nebular emission\cite{Sneppen2026}.    

The collisional process becomes more significant if the emission region is locally optically thick for H$\alpha$. Considering scattering, the critical number density is given by
\begin{equation}
    n_{\rm crit}=\frac{A_{32}}{qN_{\rm scat}},
\end{equation}
where $N_{\rm scat}$ is the number of scatterings in the local emission region. For a photon injected at line centre $N_{\rm scat}\sim\tau_{0}$, so that $\tau_{0}\sim10^{3}$--$10^{4}$ at $\NHItwo\sim10^{15.5}$--$10^{16.5}$\,\cmtwo\ places $n_{\rm crit}$ at $\sim10^{10}$--$10^{11}$\,cm$^{-3}$, within the density range of the photoionization models above.
Equivalently, a trapped \ha photon feeds the $n=3$ population rather than escaping\cite{Draine2011}. The Ly$\beta$ channel amplifies this further: each Ly$\beta$ scattering is an absorption $1s\rightarrow3p$ followed by re-emission, and is therefore itself a visit to $n=3$, while $n_2/n_1\ll1$ makes the Ly$\beta$ optical depth far exceed that of \ha. With the $3p$ branching ratio of $0.88$ into Ly$\beta$ and $0.12$ into \ha, a photon undergoes of order eight Ly$\beta$ scatterings for each conversion into \ha, so the number of $n=3$ visits exceeds the \ha scattering count above by a comparable factor, lowering $n_{\rm crit}$ to $\sim10^{9}$--$10^{10}$\,cm$^{-3}$. In addition, the population of the $ n=3$ level is enhanced by trapping, and the Paschen line also becomes optically thick.

The emission region cannot be arbitrarily optically thick, however, and the \hb line bounds it. The line-center cross-sections scale as $f\lambda$, giving $\sigma_{0}(\mathrm{H}\beta)/\sigma_{0}(\mathrm{H}\alpha)\approx0.14$, so a region with $\tau_{0,\mathrm{H}\alpha}\sim10^{3}$--$10^{4}$ has $\tau_{0,\mathrm{H}\beta}\sim10^{2}$--$10^{3}$. At such depths, \hb is efficiently removed by $4p\rightarrow3$ branching and converted to \paa and \ha \cite{ChangScattering2026}, so the region would destroy most of its own \hb. Requiring instead $\tau_{0,\mathrm{H}\beta}\lesssim10$, so that \hb survives, caps $\tau_{0,\mathrm{H}\alpha}$ at ${\sim}70$ and raises the density needed for collisional de-excitation to $n_{e}\gtrsim4\times10^{12}$\,cm$^{-3}$, above the values quoted above. The observed Balmer ratios therefore bound this channel, unless the observed \hb originates in gas other than the region invoked to suppress \ha.     

Collisions also act in the opposite direction. Collisional excitation out of $n=2$ populates $n=3$ and adds \ha beyond its recombination value, overtaking recombination once $n_2/n_p\gtrsim3\times10^{-6}$ at $T=10^{4}$\,K ($4\times10^{-5}$ at $5000$\,K), a ratio independent of density. In a gas with the enhanced $n=2$ populations inferred here, collisional excitation therefore raises the intrinsic \ha equivalent width and widens the discrepancy, while only collisional de-excitation relieves it. Modeling the balance requires solving the level populations in an optically thick medium, which the present calculation does not attempt.

Understanding the collisional processes and radiative transfer in the local emission region is therefore important for the origin of the LRD continuum and for line formation. The present work models the scattering region, but the emission region that produces the Balmer lines and the recombination continuum warrants the same treatment. If that region is locally optically thick and dense enough for collisional processes to operate, the densities $n_{\rm H}\sim10^{10}$--$10^{11}$\,cm$^{-3}$ suggested for LRDs\cite{Naidu2025,Inayoshi2025} are directly relevant to how the lines form.

\subsection*{Path length enhancement of resonance scattered H$\alpha$}

Extended Data Fig.~\ref{fig:ext_path_length} measures how much extra \hi column an \ha photon samples because of resonance scattering, and converts that into an effective $n=3$ bound--free optical depth, to test whether \ha destruction during scattering can lower the intrinsic EW established above.

We use \textsc{rt-scat}\cite{ChangScattering2026} in an idealized configuration that isolates the resonance scattering path: a central point source emitting monochromatically at \ha line center; a static outer \hi sphere ($v_{\rm exp}=0 \kms$, $\sigma_{\rm ran}=0 \kms$) at $T=10^{4}$\,K, giving a thermal Doppler parameter $b\approx12.9$\,\kms; Thomson scattering disabled ($\tau_{\rm Th}=0$); and an optically thin interclump medium. Both the homogeneous and clumpy geometries are run, for \ha and (identically) \hb, with $\NHItwo=10^{13}$--$10^{20}$\,\cmtwo\ and, for the clumpy model, covering factor $f_c\in\{1,2,5,10,20,50,100\}$ at fixed clump radius $r_{\rm cl}=10^{-3}R$. Each configuration follows $10^{5}$ photons; for \ha the escaping fraction is unity (pure scattering, no true absorption), which we monitor as a consistency check, while \hb photons are additionally removed by $4p\rightarrow3$ conversion. Because the source is monochromatic and the medium is static, this calculation isolates the longest path available under these assumptions; an intrinsic line width, bulk motion, or clump velocity dispersion would each be expected to shorten it, so the resulting enhancement is approximately an upper limit.

For each escaping photon, we accumulate the path-integrated $n=2$ column,
\begin{equation}
N_{\mathrm{HI},n=2}^{\rm path}=\sum_i n_{\mathrm{HI},n=2,i}\,\Delta s_i,
\end{equation}
summed over all propagation segments through $n=2$ gas, and define the enhancement factor
\begin{equation}
\mathcal{E}_{\rm path}\equiv\frac{\big\langle N_{\mathrm{HI},n=2}^{\rm path}\big\rangle}{\NHItwo},
\end{equation}
the ensemble mean over all escaping photons. The mean, rather than the median, is the relevant statistic here because an absorption optical depth is additive along a photon's path. Extended Data Fig.~\ref{fig:ext_path_length} plots $\mathcal{E}_{\rm path}$ against \NHItwo; the upper axis gives the \ha line-center optical depth $\tau_{0,\mathrm{H}\alpha}=\sigma_{0}\,\NHItwo$ with $\sigma_{0}\approx5\times10^{-13}$\,cm$^{2}$ at $T=10^{4}$\,K.

Across the covering factors recovered from the \ha fits ($f_c\approx3$--$90$), $\mathcal{E}_{\rm path}\approx2$--$30$ over seven decades of column in both geometries, and we adopt a representative value $\mathcal{E}_{\rm path}\sim8$ (homogeneous: $2.0$--$28.9$ over the full grid, $\approx8$ near the fitted columns; clumpy at $f_c=20$, the sample-median covering factor: $\approx3$--$8$). This is two to four orders of magnitude below the naive expectation that a resonantly trapped photon traverses of order $\tau_{0}$ radial columns before escaping ($\tau_{0,\mathrm{H}\alpha}\sim10^{4}$--$10^{6}$ at the fitted columns). The medium is instead in the regime where photons escape after frequency redistribution --- within the optically thick Doppler core at lower columns (a random walk of many short steps), or into the optically thin damping wing at higher columns (a few long flights) --- with the accumulated path growing only as $\simeq(a\tau_{0})^{1/3}$, where $a$ is the Voigt damping parameter\cite{Neufeld1990,Dijkstra2014}. In the clumpy models $\mathcal{E}_{\rm path}$ falls as $f_c$ decreases: at fixed total column, fewer clumps are each more optically thick, and a photon that strikes one is re-emitted from its thin surface layer rather than crossing it. Below $f_c\approx3$ the medium is porous --- some \ha photons escape without intersecting any clump --- and $\mathcal{E}_{\rm path}<1$.

Extended Data Fig.~\ref{fig:ext_path_length} measures path enhancement, not \ha destruction. To connect the two we form the path-integrated $n=3$ column,
\begin{equation}
N_{n=3}^{\rm path}=\frac{n_3}{n_2}\,\mathcal{E}_{\rm path}\,\NHItwo,
\end{equation}
taking for $n_3/n_2$ its LTE (Boltzmann) value at the temperature of the cool absorbing layer, computed from the level statistical weights and the $n=2$--$3$ excitation energy: $n_3/n_2\approx0.03$--$0.25$ for $T\approx5000$--$10^{4}$\,K. This is an overestimate --- resonant Ly$\alpha$ trapping enhances the $n=2$ population and lowers the true $n_3/n_2$ below its LTE value --- so the $\tau_{\rm bf,eff}$ derived below is an upper limit. The effective \ha bound--free optical depth is
\begin{equation}
\tau_{\rm bf,eff}(\mathrm{H}\alpha)=\sigma_{{\rm bf},n3}(\mathrm{H}\alpha)\,N_{n=3}^{\rm path},
\end{equation}
with $\sigma_{{\rm bf},n3}(\mathrm{H}\alpha)\approx1\times10^{-17}$\,cm$^{2}$ --- the Paschen-edge threshold cross-section ($2.1\times10^{-17}$\,cm$^{2}$ at 8206\,\AA) scaled by $(\nu_{3}/\nu_{\mathrm{H}\alpha})^{3}\approx0.5$\cite{Karzas1961}. At the fitted columns ($\NHItwo\sim10^{17}$\,\cmtwo) this gives $\tau_{\rm bf,eff}\approx0.2$--$2$; even at this LTE upper limit the \ha reduction is at most $\sim85$ percent, short of the factor $\sim10$--$100$ the EW discrepancy requires.

An \ha photon that photoionizes hydrogen from $n=3$ is not necessarily lost, since a subsequent recombination can regenerate \ha. Writing the destruction probability per $n=3$ photoionization as $\epsilon_{\rm dest}$, so that $\tau_{\rm dest,eff}\simeq\epsilon_{\rm dest}\,\tau_{\rm bf,eff}$, the estimate above takes the conservative maximum $\epsilon_{\rm dest}=1$ (every photoionization permanently removes one \ha photon). Since the reduction falls short of the required factor $\sim10$--$100$ even in this limit, allowing for recombination recycling ($\epsilon_{\rm dest}<1$) does not change the conclusion.

Producing a factor-of-$10$ \ha suppression would instead require $N_{n=3}\sim3\times10^{16}$\,\cmtwo. The same $n=3$ column implies a Paschen continuum break of optical depth $\tau\sim0.6$ and a \paa line-center optical depth $\sim10^{3}$--$10^{4}$ (the bound--bound to bound--free cross-section ratio $\sigma_{\rm bb}(\mathrm{Pa}\alpha)/\sigma_{{\rm bf},n3}\approx10^{4}$), i.e. saturated Paschen line absorption. 
Saturated Paschen-line absorption is absent in the LRDs with measured Paschen lines, even though these same objects show clear Balmer absorption \cite{Brazzini2025,Kokorev2026GLIMPSE}; the weaker continuum break is not yet constrained and could be partly filled in by Paschen recombination continuum. The same constraint applies to the collisional route above, but for a different reason: bound--free suppression requires a large $n=3$ population as a prerequisite, whereas collisional de-excitation under line trapping produces one as a consequence.  
Both therefore predict the same Paschen signature, and its depth scales with the claimed suppression, reaching $\tau_{\rm Pa\alpha}\approx0.94\,\tau_{\mathrm{H}\alpha}$ at $10^{4}$\,K in the fully thermalized limit. Paschen features lie redward of rest-frame $1\,\mu$m and are difficult to measure at these redshifts, so this is not a proof; but it shows that resolving the EW puzzle through $n=3$ bound--free suppression of \ha carries an observational cost, saturated Paschen-line absorption that is not observed, which a separate optical-continuum component does not, and that Paschen-band spectroscopy is a clean discriminating test \cite{Sneppen2026}. 

Dust along this same enhanced path could similarly extinguish \ha relative to the continuum, independent of the bound--free mechanism above. However, current LRD models generally invoke dust-poor or dust-free line-emitting gas, and direct evidence for dust in this medium is lacking, so we do not pursue this further.


\begin{appendices}

\section{Extended Data}\label{sec:extended_data}
\setcounter{figure}{0}
\renewcommand{\thefigure}{\arabic{figure}}
\setcounter{table}{0}
\renewcommand{\thetable}{\arabic{table}}

\begin{table*}[ht]
\caption{\textbf{LRD sample.} Objects sorted by redshift. $z_{\rm [OIII]}$: systemic redshift from the [O\,\textsc{iii}]$_{4960,5008}$ doublet. Grating: NIRSpec disperser used for Balmer-line profile fitting. $\mathcal{R}$: approximate spectral resolution at H$\beta$. S/N(H$\alpha$), S/N(H$\beta$): integrated-line signal-to-noise measured from our flux-calibrated spectra. PID: JWST programme ID. PI: proposal principal investigator(s).
} \label{tab:sample}
\resizebox{\linewidth}{!}{%
\begin{tabular}{lccccccc}
\hline
Object & $z_{\rm [OIII]}$ & Grating & $\mathcal{R}$ & S/N(H$\alpha$) & S/N(H$\beta$) & PID & PI \\
\hline 
RUBIES UDS 154183   & 3.548 & G235H & 2700 & 49 & 7 & 9433 & Maiolino \\
A2744 45924         & 4.464 & G395M & 1000 & 498 & 129 & 8204 & Greene \& Labbe \\
JADES GN 68797      & 5.039 & G395M & 1000 & 174 & 13 & 1181 & Eisenstein \\
GN 15498            & 5.085 & G395H & 2700 & 26 & 5 & 5664 & Matthee \\
RUBIES EGS 42046    & 5.276 & G235M & 1000 & 88 & 11 & 4233 & De Graaff \& Brammer \\
GS 13971            & 5.482 & G395H & 2700 & 57 & 10 & 5664 & Matthee \\
GN 9771             & 5.535 & G395H & 2700 & 342 & 39 & 5664 & Matthee \\
GS 3073    & 5.553 & G395H & 2700 & 242 & 98 & 1216 & L\"utzgendorf \\
JADES GN 38147 & 5.867 & G395M & 1000 & 36 & 7 & 1181 & Eisenstein \\
RUBIES EGS 49140    & 6.685 & G235M & 1000 & 57 & 17 & 4233 & De Graaff \& Brammer \\
RUBIES EGS 55604    & 6.983 & G235M & 1000 & 55 & 15 & 4233 & De Graaff \& Brammer \\
\hline
\end{tabular}
}
\end{table*}

\begin{table}
\caption{Free parameters of the Balmer-line radiative-transfer models. $N_{\rm HI,\mathit{n}=2}$ is the total \hi column density for both models; in the clumpy case, the individual clump column density is $N_{\rm HI,cl} = \tfrac{3}{4}\,N_{\rm HI,\mathit{n}=2}/f_c$ (derived). Dashes indicate the parameter does not apply to that model; `same' denotes an identical range to the homogeneous model. $T_{\rm e}$ and $r_{\rm cl}$ are fixed.}
\label{tab:rt_params}
\begin{tabular}{lll}
\hline
Parameter & Homogeneous & Clumpy \\
\hline
$N_{\rm HI,\mathit{n}=2}$ [cm$^{-2}$] & $10^{12}$--$10^{20}$ & same \\
$v_{\rm exp}$ [\kms] & $-400$ to $+400$ & same \\
$\sigma_{\rm emit}$ [\kms] & $100$--$300$ & same \\
$\sigma_{\rm ran}$ [\kms] & $50$--$150$ & same \\
$\tau_{\rm Th}$ & $0.5$--$3.0$ & same \\
$f_c$ & --- & $1$--$100$ \\
\hline
\end{tabular}
\end{table}

\begin{table*}
\caption{\textbf{Balmer-break SED fitting results for all 11 LRDs.} $\alpha_{\rm UV}$: UV power-law slope from Stage~1. $T_{\rm BB}$: effective optical continuum temperature from Stage~2; this is a fitting parameter, not a physical temperature (see text). $f_{\rm exc}$: optical-excess parameter (equation~\ref{eq:fexc}); these three columns do not depend on the H\,\textsc{i} velocity model. $\log\,N_{\rm HI,\mathit{n}=2}$ and $v_{\rm ran}$ are the Stage-3 high-n Balmer series fit values (see text), reported separately for the clumpy and homogeneous models since each is fit with its own model-specific H$\alpha$-derived prior on $\log\,N_{\rm HI,\mathit{n}=2}$. Superscripts/subscripts give the 84th/16th percentiles of the Stage-1, Stage-2, and Stage-3 posteriors. The last two rows (GS~3073 and JADES~GN~38147) did not satisfy the detection criterion ($f_{\rm exc} < 1$); dashes indicate that the Stage-3 high-n Balmer series fit is not meaningful for these objects. For JADES~GN~38147, the optical excess is too weak to constrain $T_{\rm BB}$.
$\ddagger$: the 84th percentile of \NHItwo is below $10^{16.5}$\cmtwo ($\tau_{\rm BB}=0.5$), too low to produce the Balmer break; the median is given and $v_{\rm ran}$ is unconstrained.}
\label{tab:sed_fits}
\resizebox{\linewidth}{!}{%
\begin{tabular}{lccccccccc}
\hline
& & & & & \multicolumn{2}{c}{Clumpy} & \multicolumn{2}{c}{Homogeneous} \\
Object & $z$ & $\alpha_{\rm UV}$ & $T_{\rm BB}$ (K) & $f_{\rm exc}$ & $\log\,N_{\rm HI,\mathit{n}=2}$ (cm$^{-2}$) & $v_{\rm ran}$ (\kms) & $\log\,N_{\rm HI,\mathit{n}=2}$ (cm$^{-2}$) & $v_{\rm ran}$ (\kms)  \\
\hline
GN 9771 & 5.535 & $-0.87^{+0.05}_{-0.05}$ & $5344^{+114}_{-109}$ & $1.95^{+0.13}_{-0.13}$ & $17.02^{+0.03}_{-0.03}$ & 
$400^{+67}_{-85}$ & $13.75$$^\ddagger$ & --- \\
A2744 45924 & 4.464 & $-0.43^{+0.06}_{-0.06}$ & $4802^{+71}_{-70}$ & $2.65^{+0.19}_{-0.19}$ & $17.20^{+0.06}_{-0.06}$ & 
$388^{+74}_{-88}$ & $14.08$$^\ddagger$ & --- \\
RUBIES EGS 49140 & 6.685 & $-0.62^{+0.12}_{-0.11}$ & $4712^{+224}_{-209}$ & $2.80^{+0.31}_{-0.27}$ &
$18.29^{+0.38}_{-0.46}$ & $275^{+96}_{-80}$ & $17.86^{+0.44}_{-0.32}$ & $300^{+100}_{-87}$ \\
RUBIES EGS 55604 & 6.983 & $-1.08^{+0.10}_{-0.10}$ & $4644^{+329}_{-300}$ & $2.27^{+0.23}_{-0.24}$ &
$17.54^{+0.24}_{-0.19}$ & $356^{+89}_{-94}$ & $17.40^{+0.17}_{-0.13}$ & $367^{+84}_{-93}$ \\
JADES GN 68797 & 5.039 & $-0.70^{+0.18}_{-0.17}$ & $3396^{+89}_{-87}$ & $5.64^{+1.04}_{-0.90}$ & $17.57^{+0.25}_{-0.26}$ &
$151^{+129}_{-74}$ & $14.41$$^\ddagger$ & --- \\
RUBIES EGS 42046 & 5.276 & $-0.25^{+0.16}_{-0.16}$ & $4913^{+343}_{-312}$ & $1.26^{+0.34}_{-0.28}$ &
$17.46^{+0.30}_{-0.21}$ & $342^{+105}_{-118}$ & $17.70^{+0.28}_{-0.28}$ & $330^{+111}_{-116}$ \\
RUBIES UDS 154183 & 3.548 & $-1.13^{+0.66}_{-0.61}$ & $4106^{+90}_{-87}$ & $18.15^{+14.32}_{-7.70}$ &
$17.59^{+0.54}_{-0.34}$ & $211^{+129}_{-82}$ & $13.93$$^\ddagger$ & --- \\
GN 15498 & 5.085 & $-1.80^{+0.20}_{-0.20}$ & $4255^{+196}_{-187}$ & $6.63^{+1.68}_{-1.36}$ & $18.18^{+0.48}_{-0.58}$ &
$269^{+142}_{-131}$ & $17.92^{+0.56}_{-0.48}$ & $278^{+138}_{-135}$ \\
GS 13971 & 5.482 & $-1.65^{+0.11}_{-0.11}$ & $3680^{+251}_{-236}$ & $2.90^{+0.43}_{-0.40}$ & $16.81^{+0.47}_{-1.44}$ &
$200^{+188}_{-138}$ & $14.81$$^\ddagger$ & --- \\
\textit{GS 3073} & \textit{5.553} & $-1.61^{+0.02}_{-0.02}$ & $8699^{+1565}_{-1075}$ & $0.36^{+0.02}_{-0.02}$ & --- & ---
& --- & --- \\
\textit{JADES GN 38147} & \textit{5.867} & $-1.52^{+0.11}_{-0.11}$ & --- & $0.63^{+0.17}_{-0.16}$ & --- & --- & --- & --- 
\\
\hline
\end{tabular}}
\end{table*}

\begin{figure*}[ht]
    \centering
    \includegraphics[width=\textwidth]{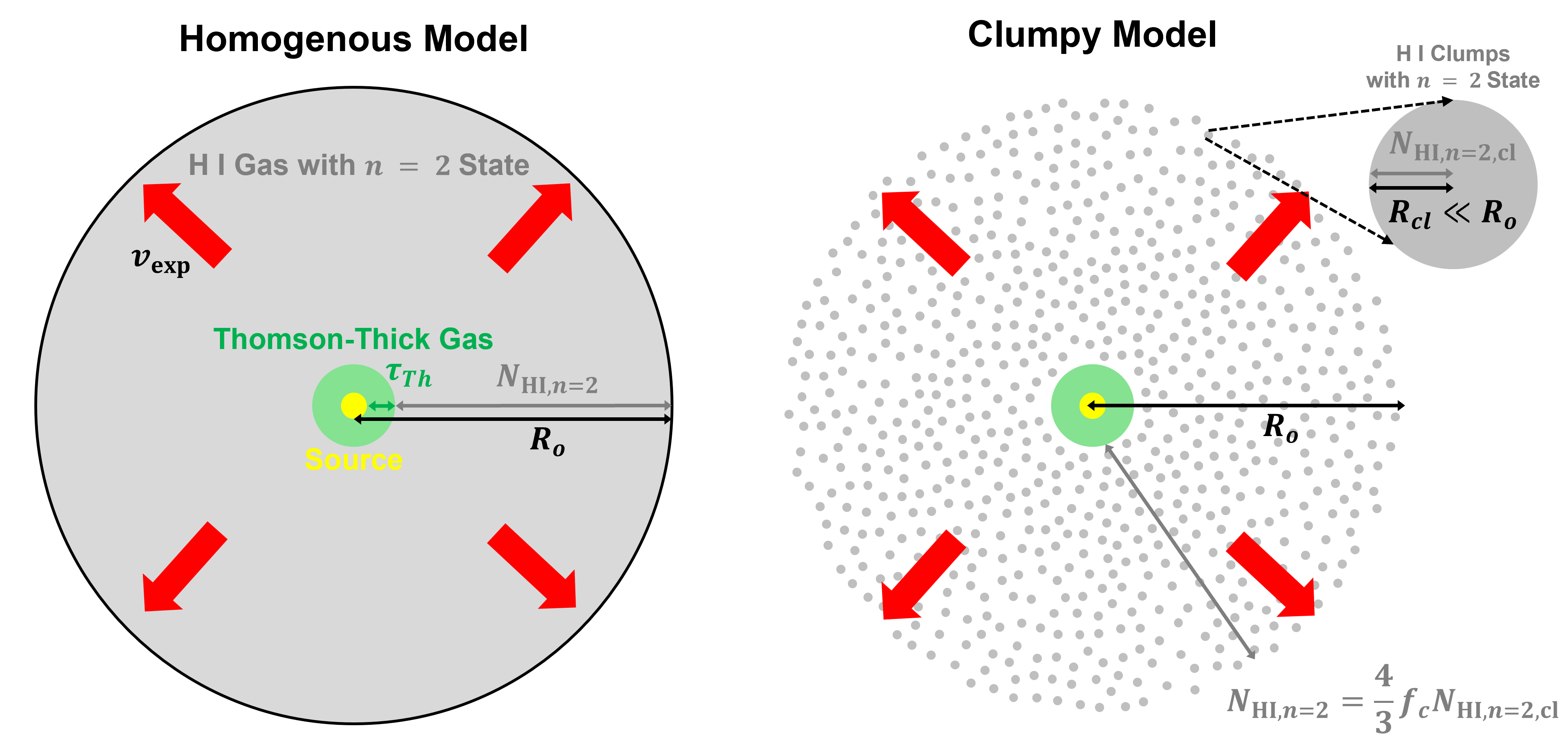}
        \caption{\textbf{Scattering geometry of the homogeneous (left) and clumpy (right) H I gas.} A central isotropic source (yellow) emits H$\alpha$ and H$\beta$ emission with a Gaussian profile of width $\sigma_{\rm emit}$.
        The source is surrounded by two concentric regions: an inner, fully ionized H II region (green, $r < 0.01\,R_o$) with Thomson optical depth $\tau_{\rm Th}$, and an outer neutral H~I region (grey, $0.01\,R_o < r < R_o$) with a significantly populated $n=2$ level and total column density $N_{\rm HI,\mathit{n}=2}$ and a random motion $\sigma_{\rm ran}$. Red arrows: bulk radial outflow velocity $v_{\rm exp}$, constant with radius ($v(r) = v_{\rm exp}$). Left: the H\,\textsc{i} region is a uniform-density sphere; this homogeneous-model geometry follows \cite{ChangScattering2026}, Fig.~15. Right: the same total column $N_{\rm HI,\mathit{n}=2}$ is instead distributed among discrete, optically thick clumps of radius $R_{\rm cl} \ll R_o$ and column density $N_{\rm HI,\mathit{n}=2,cl}$, related by $N_{\rm HI,\mathit{n}=2} = \tfrac{4}{3} f_c N_{\rm HI,\mathit{n}=2,cl}$, where $f_c$ is the covering factor, which is the average number of clumps from the central source. Individual clumps additionally carry a random peculiar velocity dispersion $\sigma_{\rm ran}$. The random motion within clumps is the thermal motion, as their temperature is at $T=10^4\, \rm K$. This clumpy-medium treatment follows \cite{SJC2023,Chang2024MgII}.}
    \label{fig:ext_geometry}
\end{figure*}

\begin{figure*}[ht]
  \centering
  \includegraphics[width=\textwidth]{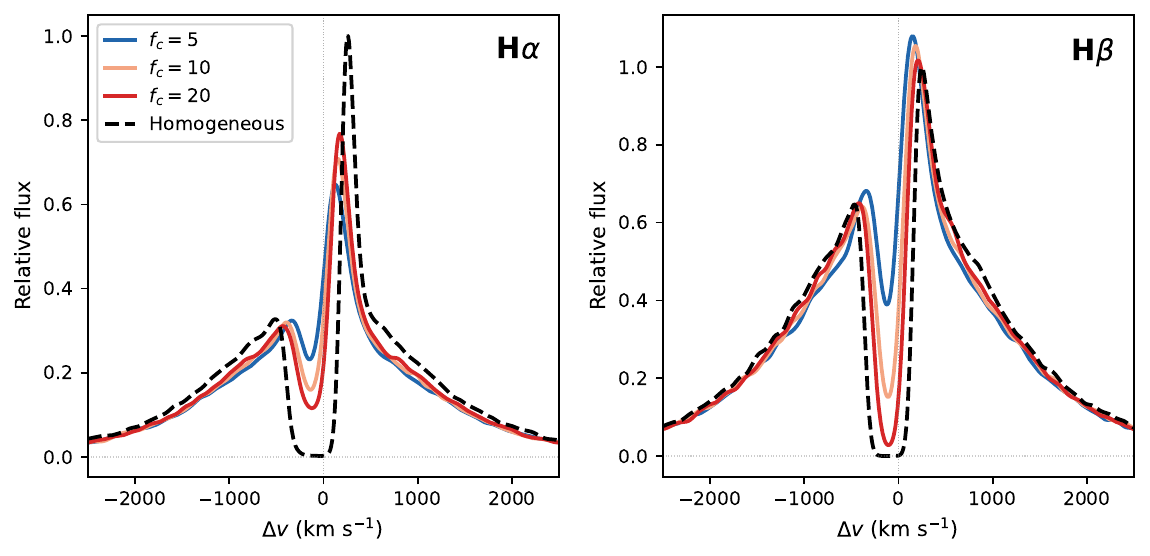}
  \caption{\textbf{Effect of covering factor $f_c$ on H$\alpha$ and H$\beta$ profiles at fixed total column density.}
  Simulated H$\alpha$ (left) and H$\beta$ (right) profiles from direct RT-SCAT Monte Carlo outputs (no emulator) at fixed representative parameters ($\log\,N_{\rm HI,\mathit{n}=2} = 17\, [\rm cm^{-2}]$, $v_{\rm exp} = 100$\,\kms, $\sigma_{\rm emit} = 200$\,\kms, $\sigma_{\rm ran} = 100$\,\kms, $\tau_{\rm Th} = 1.5$), for clumpy models with $f_c = 5$, 10, 20 and the homogeneous case. The primary contrast is between clumpy and homogeneous: all three clumpy models show dramatically weaker H$\alpha$ absorption than the homogeneous case at the same \NHItwo, while H$\beta$ absorption is reduced far less. This asymmetric response --- the porous clump geometry selectively suppresses H$\alpha$ while retaining substantial H$\beta$ absorption --- may be the physical origin of the H$\alpha$--H$\beta$ absorption discrepancy observed in LRDs. The three $f_c$ values cluster together in both panels, indicating that the surface-scattering regime is reached robustly across the range of covering factors inferred from the data\cite{Gronke2016,Gronke2017}.}
  \label{fig:ext_fc}
\end{figure*}

\begin{figure*}[ht]
  \centering
  \includegraphics[width=\textwidth]{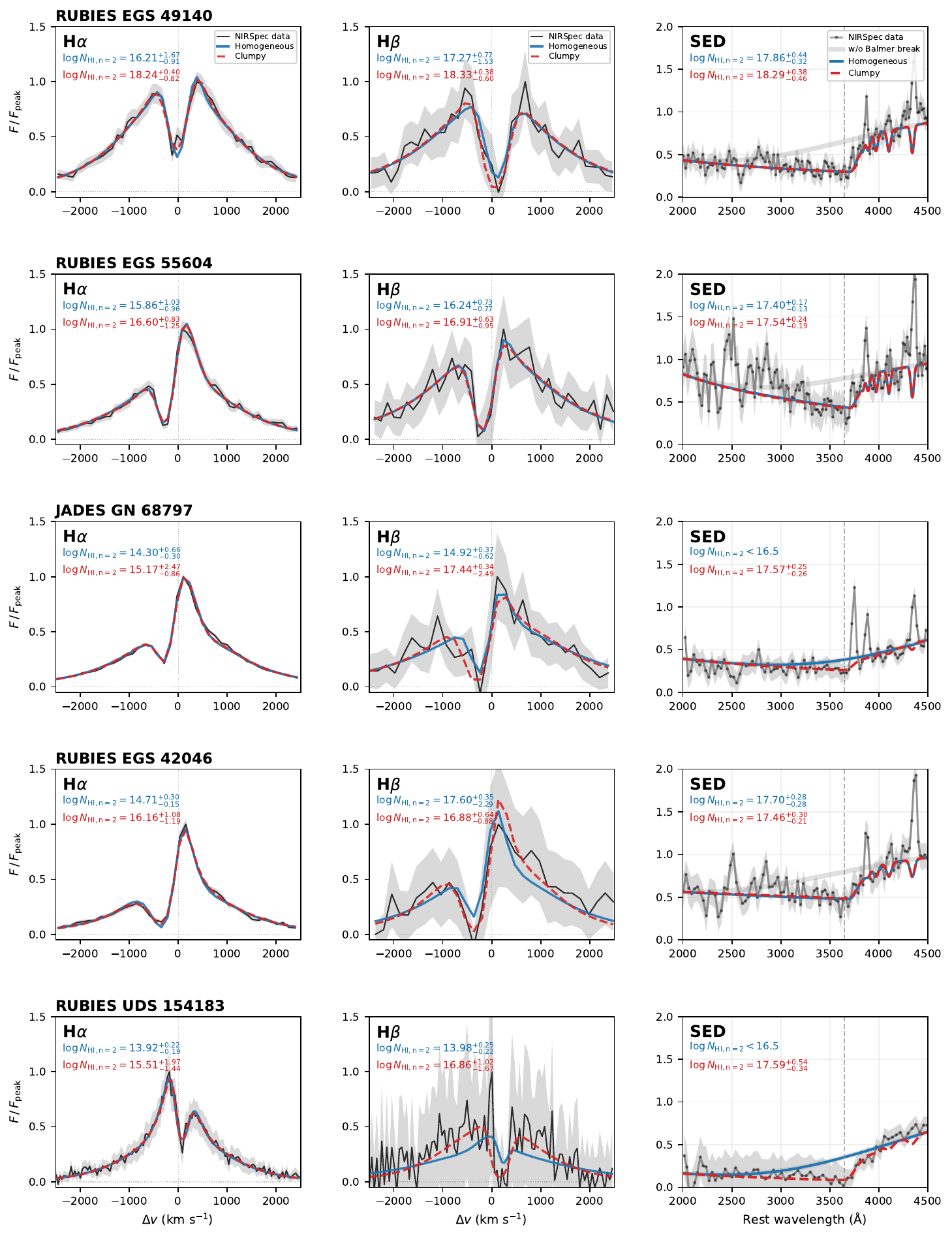}
  \caption{\textbf{H$\alpha$, H$\beta$, and Balmer-break fits for the remaining 9 LRDs
  (1 of 2).} Format and colors as in Fig.~\ref{fig:example_fit}; GN~9771 and A2744~45924 are shown in Fig.~\ref{fig:example_fit}. One row per object, in order: RUBIES~EGS~49140, RUBIES~EGS~55604, JADES~GN~68797, RUBIES~EGS~42046, and RUBIES~UDS~154183. The remaining 4 objects are shown in Extended Data Fig.~\ref{fig:ext_fit2}.}
  \label{fig:ext_fit1}
\end{figure*}

\begin{figure*}[ht]
  \centering
  \includegraphics[width=\textwidth]{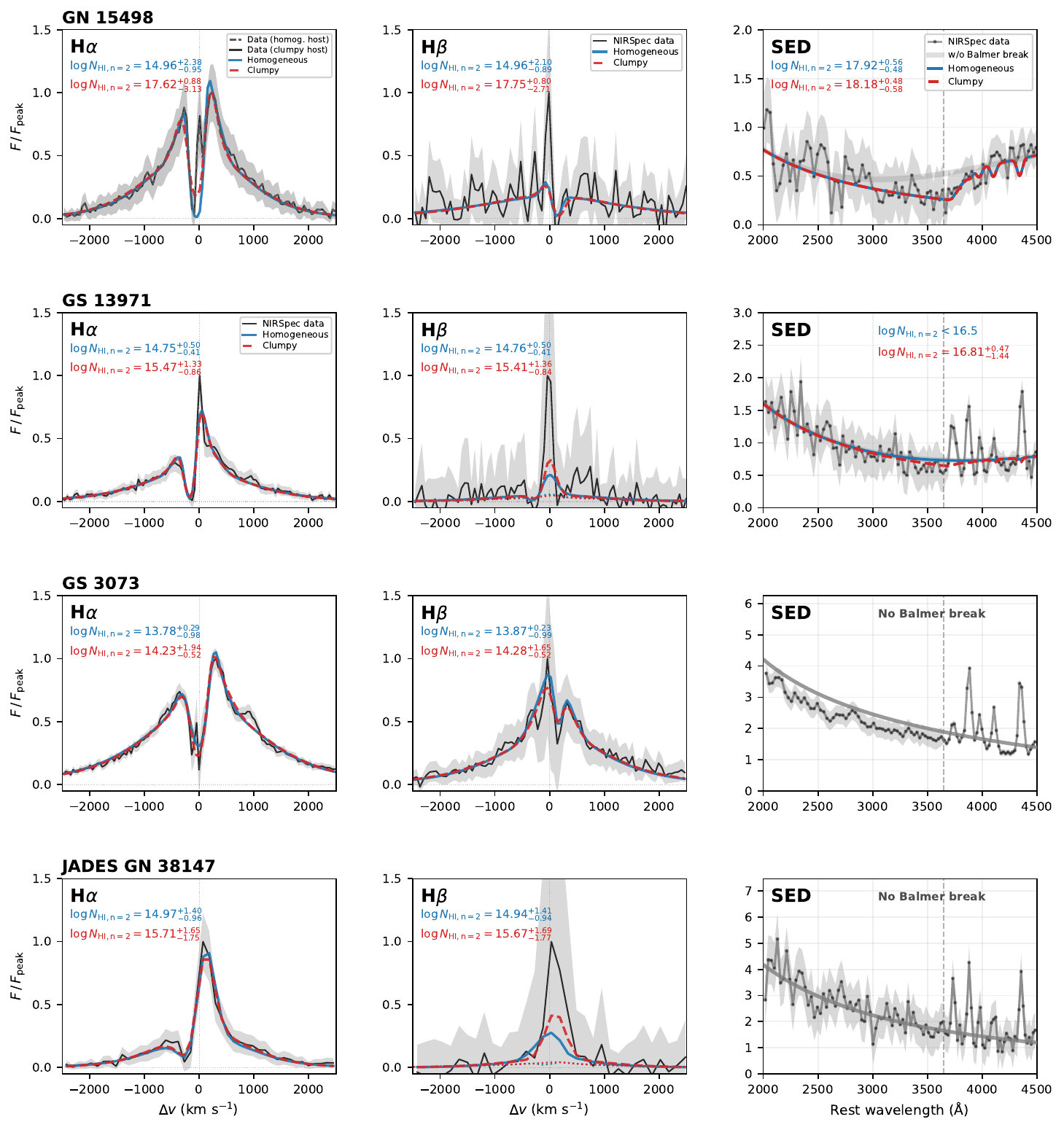}
  \caption{\textbf{H$\alpha$, H$\beta$, and Balmer-break fits for the remaining 9 LRDs
  (2 of 2).} Format and colors as in Fig.~\ref{fig:example_fit}; the first 5 objects are shown in Extended Data Fig.~\ref{fig:ext_fit1}. One row per object, in order: GN~15498, GS~13971, GS~3073, and JADES~GN~38147. GS~3073 and JADES~GN~38147 did not satisfy the Balmer-break detection criterion ($f_{\rm exc} < 1$); their SED panels show only the continuum without a break and are labeled `No Balmer Break'. For GN~15498, the H$\alpha$ panel shows both observed spectra for the clumpy and homogeneous models, since the two diverge visibly for this object due to different [O III]-based subtraction.}
  \label{fig:ext_fit2}
\end{figure*}

\begin{figure*}[ht]
  \centering
  \includegraphics[width=\textwidth]{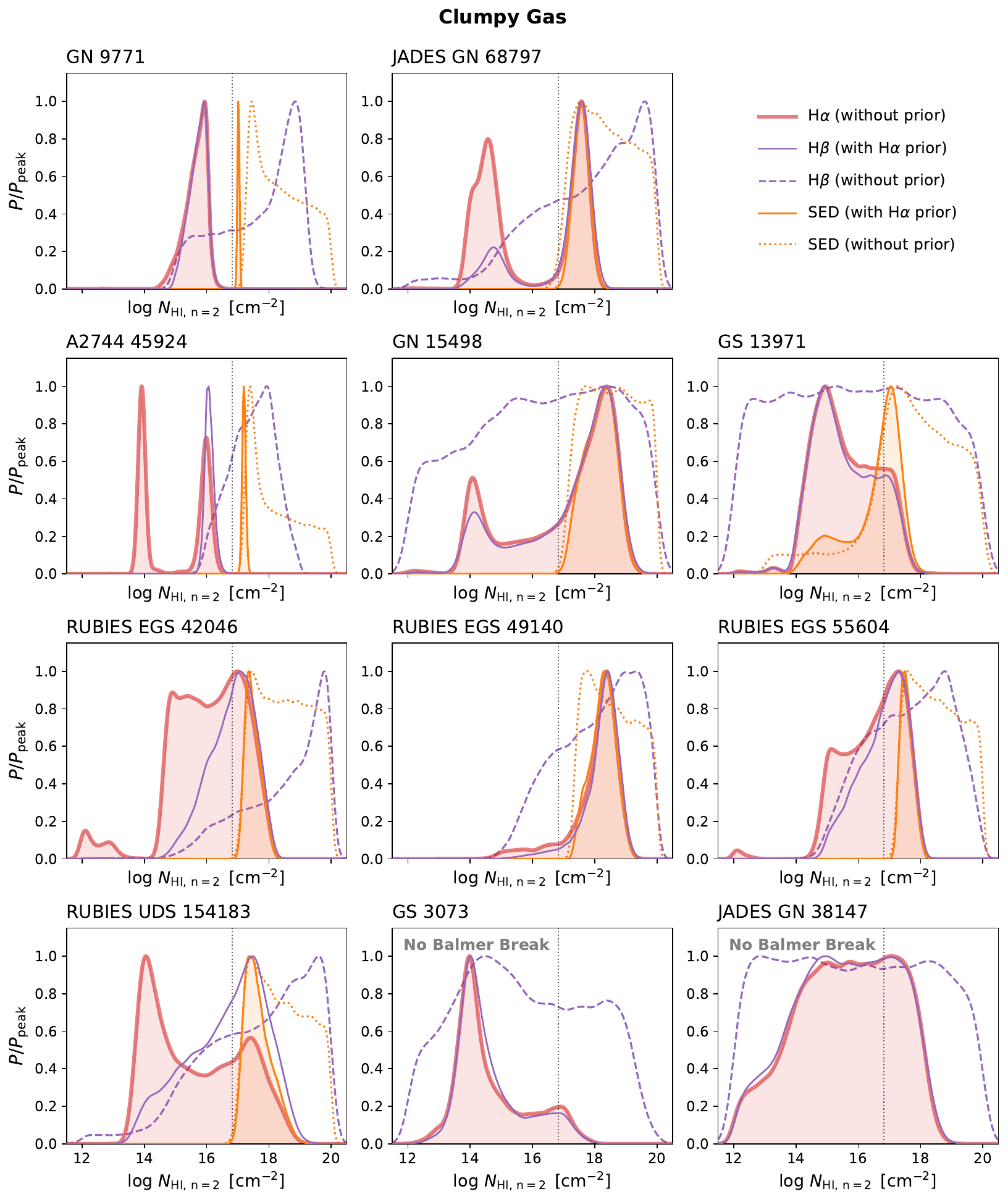}
  \caption{\textbf{$N_{\rm HI,\mathit{n}=2}$ posteriors for all 11 LRDs, clumpy model.} Each panel shows, for one object, the one-dimensional posterior probability density (peak-normalised) for $\log\,N_{\rm HI,\mathit{n}=2}$ from five probes: the H$\alpha$-only fit (solid); the H$\beta$ fit with the H$\alpha$-derived \NHItwo prior (solid) and without this prior (dashed); and the Balmer-break SED fit with the H$\alpha$-derived prior (solid) and without this prior (dotted). The H$\alpha$ and SED-with-prior posteriors are shaded, and the vertical dotted line marks $\tau_{\rm BB}=1$. H$\alpha$ and H$\beta$ posteriors are from the full MCMC chains; SED posteriors are from the Stage-3 MCMC chains (see Methods). GS~3073 and JADES~GN~38147 have no SED posteriors because these two objects have no Balmer break.}
  \label{fig:ext_nhi_post_clumpy}
\end{figure*}

\begin{figure*}[ht]
  \centering
  \includegraphics[width=\textwidth]{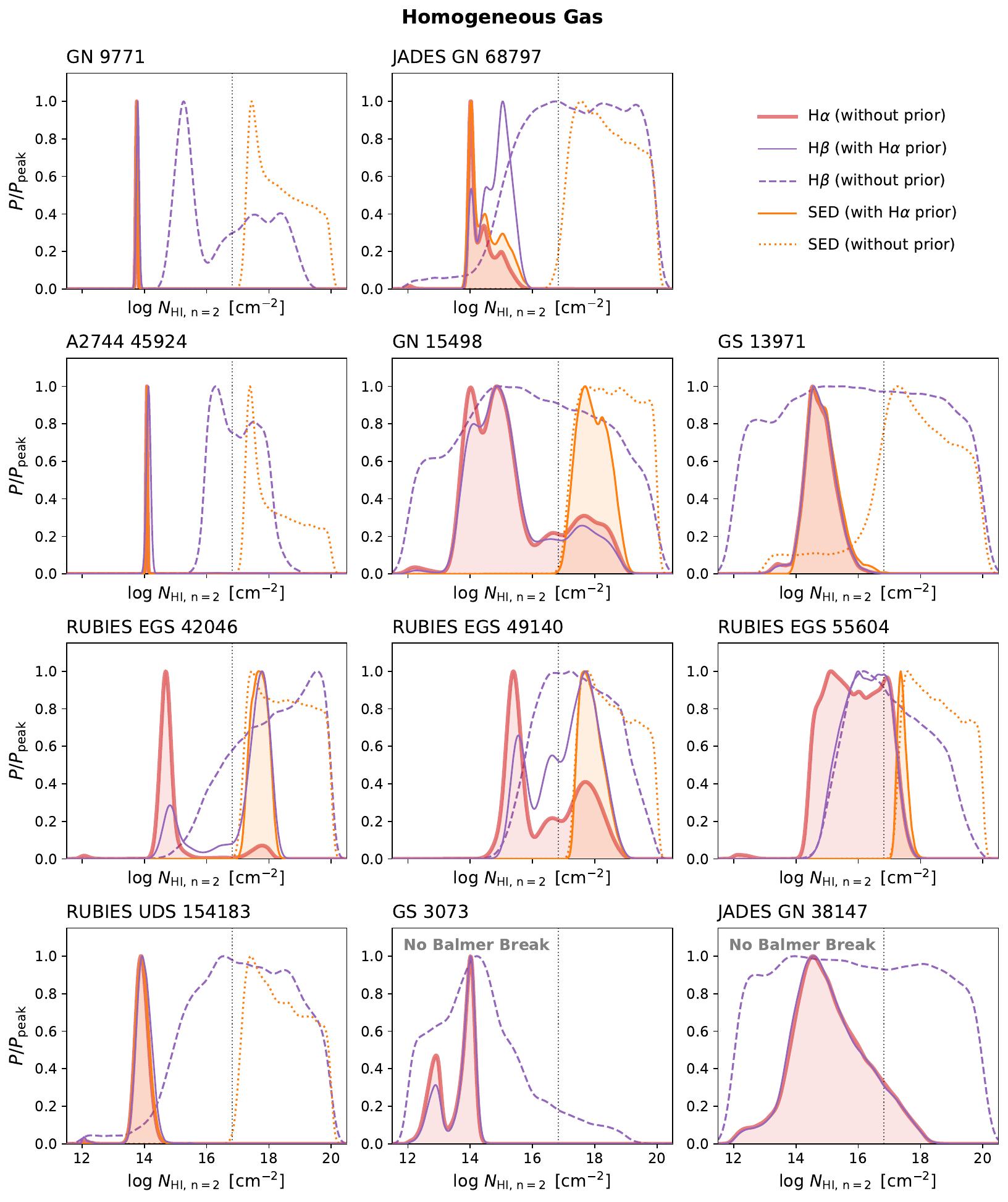}
  \caption{\textbf{$N_{\rm HI,\mathit{n}=2}$ posteriors for all 11 LRDs, homogeneous model.} As Extended Data Fig.~\ref{fig:ext_nhi_post_clumpy}, for the homogeneous radiative-transfer model.}
  \label{fig:ext_nhi_post_smooth}
\end{figure*}

\begin{figure*}[ht]
  \centering
  \includegraphics[width=\textwidth]{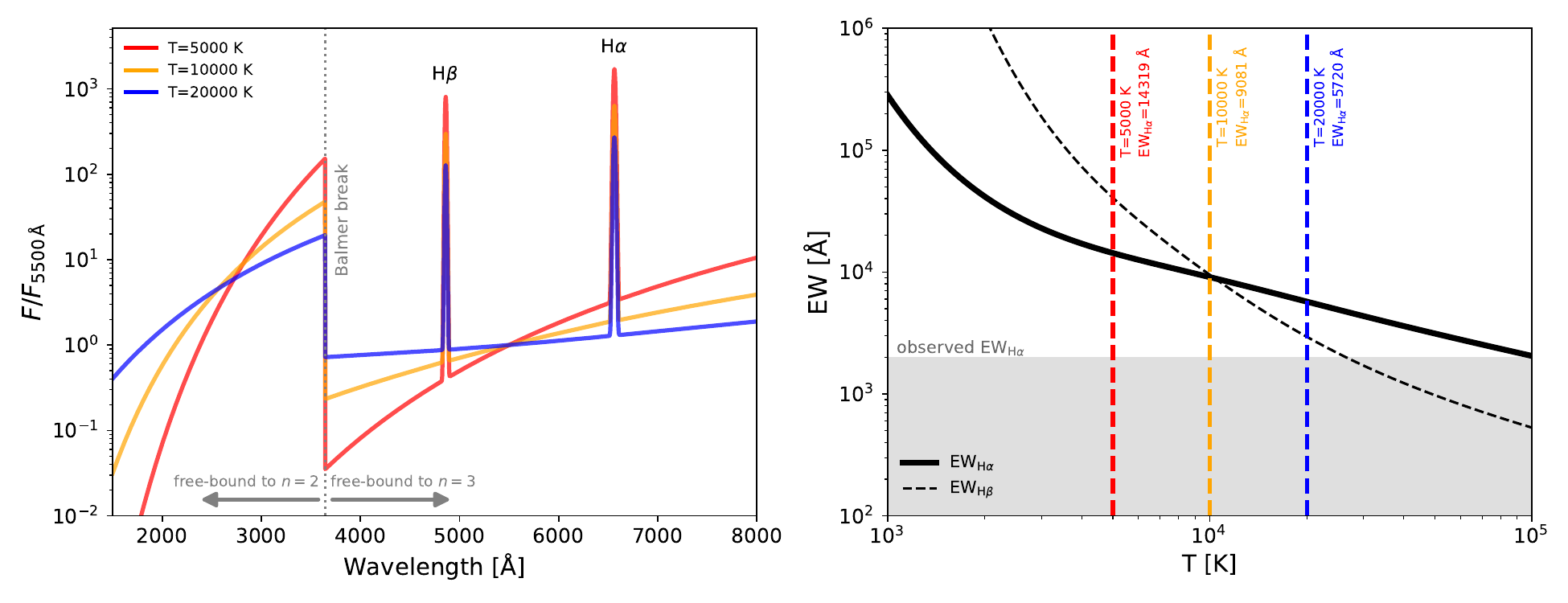}
  \caption{\textbf{Nebular continuum with H$\alpha$ and H$\beta$ emission, and the equivalent-width cost of a nebular-dominated optical continuum.} \textit{Left:} the hydrogen nebular recombination continuum (free--free $+$ free--bound; two-photon emission omitted) at $T=5000$, $10^{4}$, and $2\times10^{4}$\,K, normalized to unity at rest-frame 5500\,\AA, with the \ha and \hb emission it implies drawn on top at the self-consistent single-zone Case~B equivalent widths of the right panel (Gaussian, $500\,\mathrm{km\,s^{-1}}$). \textit{Right:} the intrinsic $\rm{EW}_{\mathrm{H}\alpha}$ implied if the nebular continuum alone supplies the observed optical continuum, versus $T$ (electron density and emitting volume cancel in this ratio); vertical lines mark the three left-panel temperatures, the shaded band the representative observed LRD range, $\sim100$--$2000$\,\AA. The thin dashed curve shows the corresponding $\rm{EW}_{\rm{H}\beta}$, shown for consistency with the H$\beta$ line drawn in the left panel and not as an independent constraint. Over the plotted range $\rm{EW}_{\rm{H}\alpha}\sim10^{3}$--$10^{5}$\,\AA, one to two orders of magnitude above typical observed LRD values. See Methods.}
  \label{fig:ext_neb_bb_ew}
\end{figure*}

\begin{figure*}[ht]
  \centering
  \includegraphics[width=\textwidth]{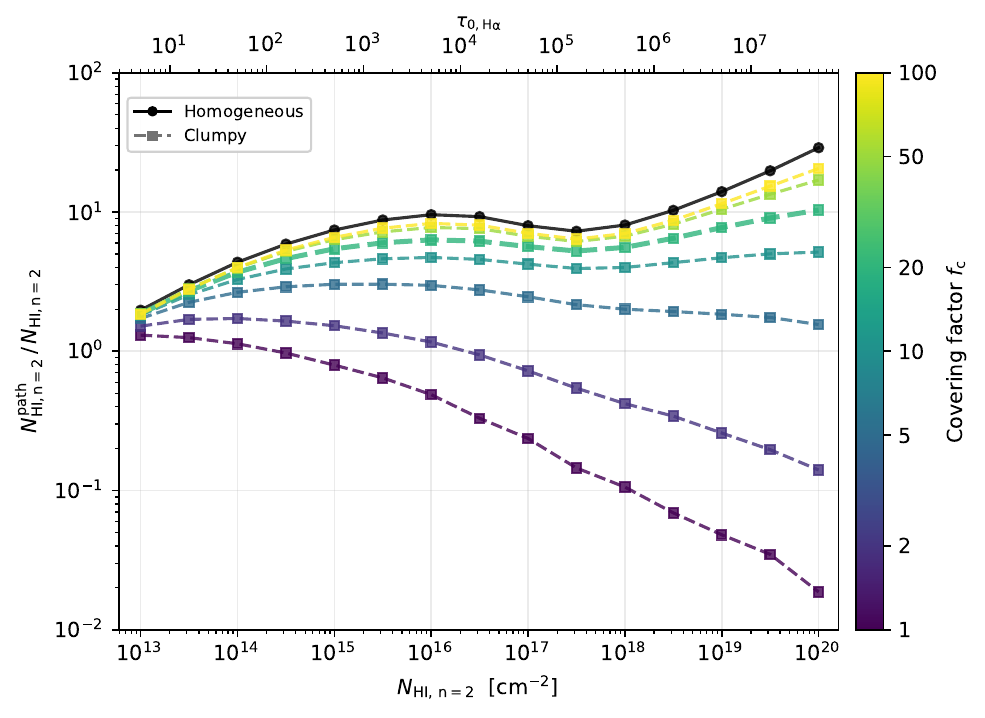}
  \caption{\textbf{The $n=2$ neutral-hydrogen column sampled by a resonantly scattered \ha photon exceeds the radial column by only a factor $\sim$$8$.} Ensemble-mean path-integrated $n=2$ column divided by the radial column \NHItwo, from \textsc{rt-scat} Monte Carlo runs of $10^{5}$ line-center \ha photons per grid point through a static \hi medium ($\tau_{\rm Th}=0$), for the homogeneous model (solid) and clumpy models with $r_{\rm cl}=10^{-3}R$ and covering factor $f_c=1$--$100$ (dashed, color-coded; $f_c=20$, the sample-median value, drawn heavier). Upper axis: \ha line-center optical depth $\tau_{0,\mathrm{H}\alpha}=\sigma_{0}\,\NHItwo$, $\sigma_{0}\simeq5\times10^{-13}\,\mathrm{cm^{2}}$. The amplification is $\approx2$--$30$ across seven decades of column in both geometries (representative value $\sim$$8$) --- two to four orders of magnitude below the $\sim\tau_{0}$ expected for a core-trapped photon. Converted to an $n=3$ bound--free optical depth, this is short of the factor $\sim10$--$100$ suppression of \ha the equivalent-width puzzle would require (see Methods). \hb (not shown): ratio below unity throughout.}
  \label{fig:ext_path_length}
\end{figure*}

\begin{figure*}[ht]
  \centering
  \includegraphics[width=\textwidth]{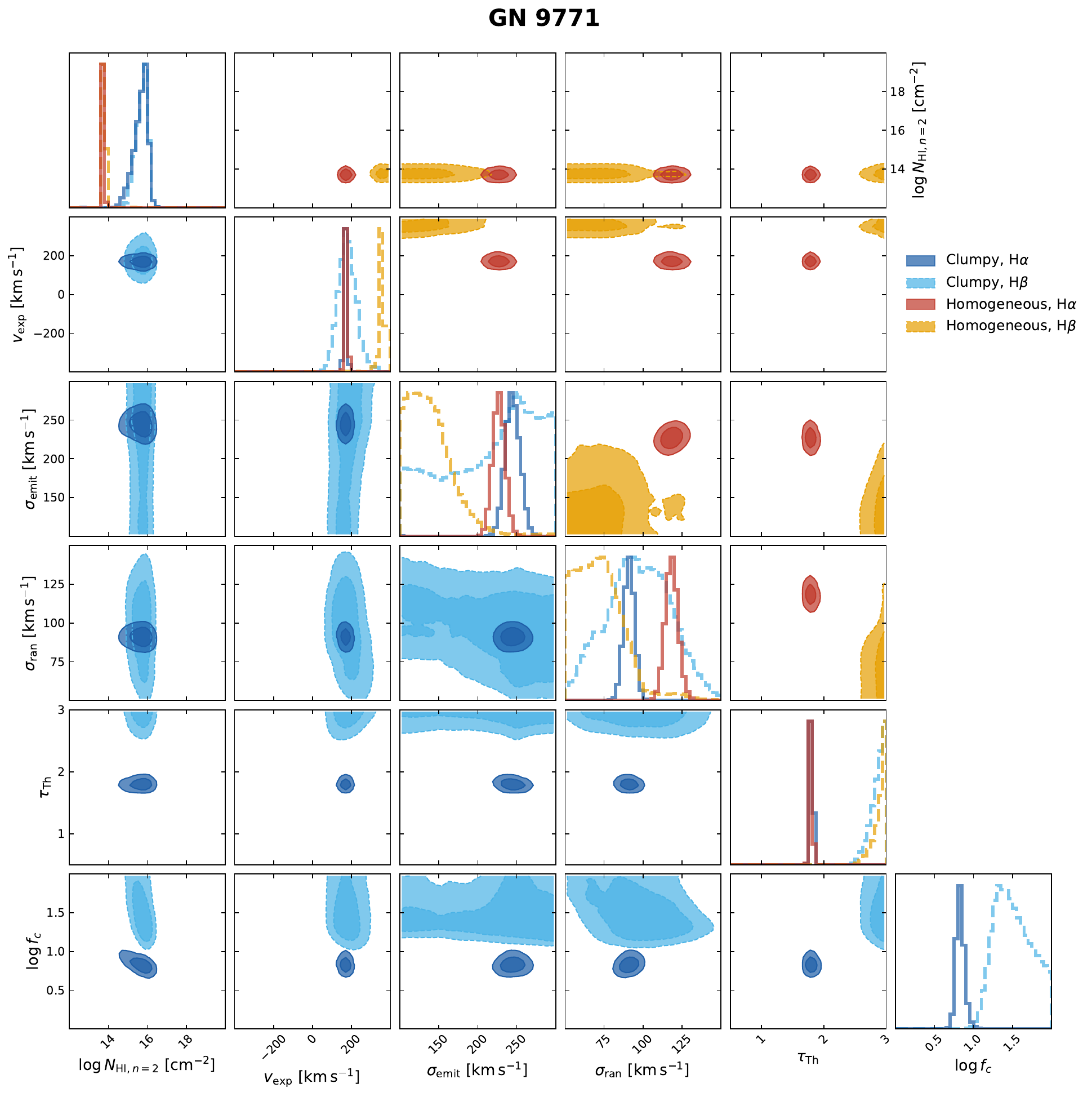}
  \caption{\textbf{Posterior distributions of the Balmer-line radiative-transfer fits for GN~9771.}
  The lower-left triangle shows the clumpy model and the upper-right triangle the homogeneous model, with the axes of the upper triangle transposed. 
  Dark blue and red show the H$\alpha$-only fit (Stage~1), and light blue and orange show the H$\beta$-only fit (Stage~2), which uses the \NHItwo prior from the H$\alpha$ fit of the same model (see Methods).
  Filled contours mark the 68 and 95 percent credible regions; H$\beta$ contours have dashed outlines. 
  Diagonal panels show the one-dimensional marginalized posteriors, each normalized to unit peak (solid: H$\alpha$, dashed: H$\beta$).
  The homogeneous model has no $f_c$, so its column is left empty.
  Axis ranges span the full prior ranges of Extended Data Table~\ref{tab:rt_params}; the remaining nuisance parameters are marginalized over.
  With the H$\alpha$-based prior, the H$\beta$ fits of both models stay at \NHItwo from H$\alpha$ ($\log \NHItwo \approx 15.8$ and 13.8 $[\rm cm^{-2}]$ for clumpy and homogeneous). The homogeneous H$\beta$ fit is pushed to the upper prior bounds of $v_{\rm exp}$ and $\tau_{\rm Th}$ and still underpredicts the H$\beta$ absorption (Fig.~\ref{fig:example_fit}), whereas the clumpy H$\beta$ fit reproduces it.}                    
  \label{fig:ext_corner_gn9771} 
\end{figure*}

\begin{figure*}[ht]
  \centering
  \includegraphics[width=\textwidth]{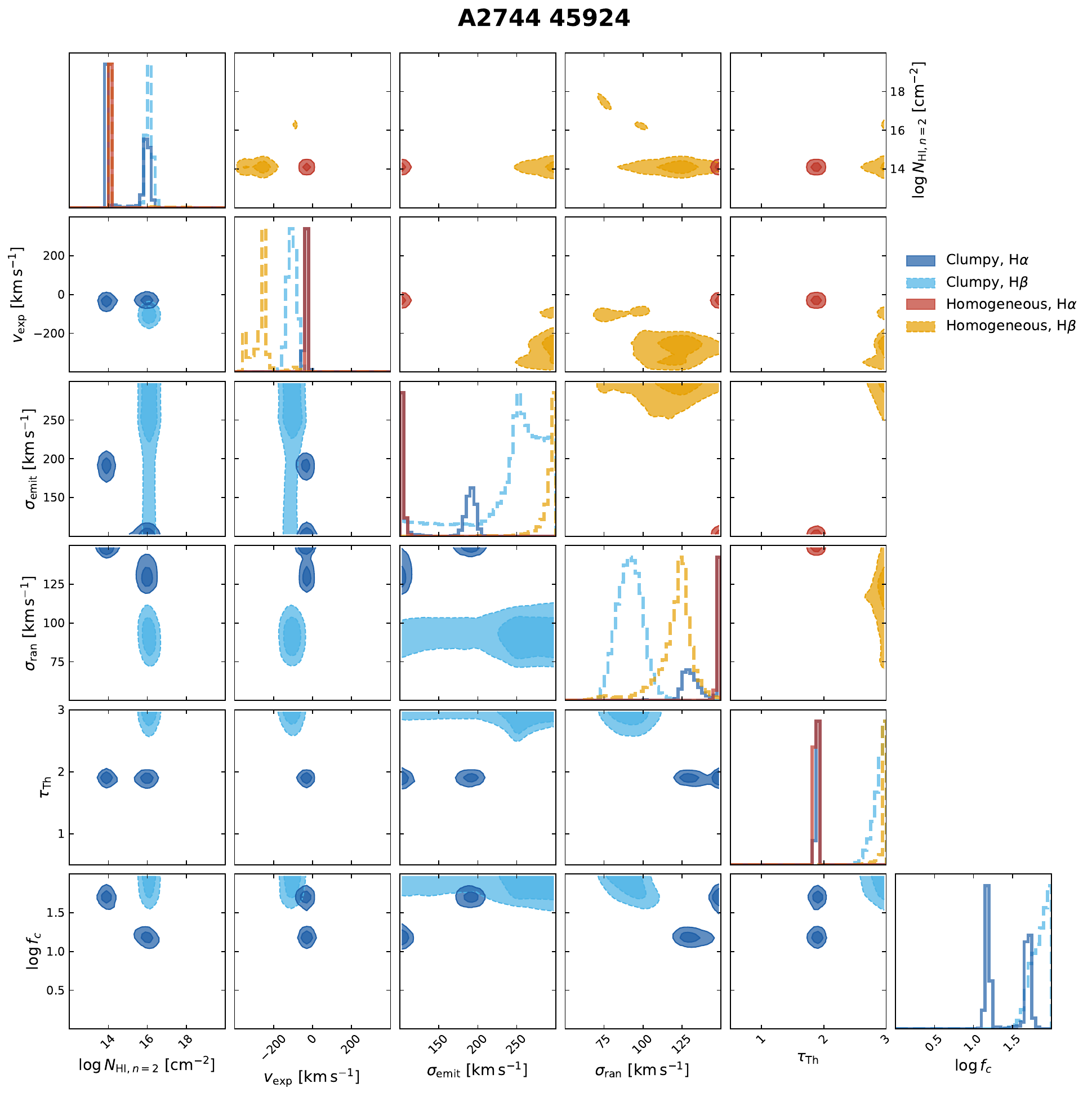}
  \caption{\textbf{Posterior distributions of the Balmer-line radiative-transfer fits for A2744~45924.}
  Same as Extended Data Fig.~\ref{fig:ext_corner_gn9771}.
  The clumpy H$\alpha$ posterior is bimodal, with modes at $\log \NHItwo \approx 14$ and 16 $[\rm cm^{-2}]$ that differ in $f_c$; the clumpy H$\beta$ fit selects the upper mode ($\log \NHItwo \approx 16.1$). The homogeneous H$\beta$ fit stays close to \NHItwo from H$\alpha$ ($\log \NHItwo \approx 14.1$), is pushed to the upper $\tau_{\rm Th}$ and $\sigma_{\rm emit}$ bounds, and still underpredicts the H$\beta$ absorption (Fig.~\ref{fig:example_fit}). For this object, the host Balmer decrement sits at its upper limit of 3.3 in both models.}
  \label{fig:ext_corner_a2744}
\end{figure*}

\begin{table*}
\caption{\textbf{Homogeneous (smooth) model fitted parameters.}
Median and 68 percent credible intervals for all 11 LRDs, from the
individual-fit chain (see Methods): for each object, the H$\alpha$ row
is the H$\alpha$-only fit; the H$\beta$ row is the H$\beta$ fit using
the \NHItwo posterior of this H$\alpha$ fit as a prior. 
Object order as in Fig.~\ref{fig:nhi_comparison}. Italicized rows (GS~3073 and
JADES~GN~38147) have no Balmer-break detection; line-fitting results are
included for completeness. $v_{\rm exp}$: envelope expansion velocity
(negative values indicate infall). $\sigma_{\rm emit}$: intrinsic emission-line width.
$\sigma_{\rm ran}$: random velocity dispersion. $\tau_{\rm Th}$: Thomson
optical depth. $\chi^2_{\rm r}$: $\chi^2$ of the median posterior model per
spectral bin ($|v| < 3000$\,\kms for H$\alpha$, $< 2000$\,\kms for H$\beta$).}
\label{tab:ext_smooth}
\resizebox{\linewidth}{!}{%
\begin{tabular}{llcccccc}
\hline
Object & Fit
  & $\log\,N_{\rm HI,\mathit{n}=2}$ (cm$^{-2}$)
  & $v_{\rm exp}$ (km\,s$^{-1}$)
  & $\sigma_{\rm emit}$ (km\,s$^{-1}$)
  & $\sigma_{\rm ran}$ (km\,s$^{-1}$)
  & $\tau_{\rm Th}$
  & $\chi^2_{\rm r}$ \\
\hline
\multirow{2}{*}{GN~9771} & H$\alpha$ & $13.75^{+0.03}_{-0.03}$ & $173^{+6}_{-6}$ & $227^{+8}_{-8}$ & $118^{+4}_{-4}$ & $1.79^{+0.02}_{-0.02}$ & $1.30$ \\
 & H$\beta$ & $13.79^{+0.03}_{-0.03}$ & $353^{+28}_{-8}$ & $135^{+30}_{-24}$ & $71^{+15}_{-14}$ & $2.93^{+0.05}_{-0.13}$ & $1.25$ \\
\hline
\multirow{2}{*}{A2744~45924} & H$\alpha$ & $14.08^{+0.02}_{-0.02}$ & $-25^{+2}_{-2}$ & $102^{+3}_{-1}$ & $149^{+0}_{-1}$ & $1.88^{+0.01}_{-0.01}$ & $14.07$ \\
 & H$\beta$ & $14.14^{+0.02}_{-0.02}$ & $-256^{+12}_{-70}$ & $295^{+4}_{-10}$ & $123^{+6}_{-12}$ & $2.99^{+0.01}_{-0.04}$ & $5.38$ \\
\hline
\multirow{2}{*}{RUBIES~EGS~49140} & H$\alpha$ & $16.20^{+1.67}_{-0.91}$ & $12^{+8}_{-7}$ & $157^{+101}_{-44}$ & $102^{+42}_{-43}$ & $1.81^{+0.14}_{-0.16}$ & $0.49$ \\
 & H$\beta$ & $17.27^{+0.77}_{-1.53}$ & $-73^{+82}_{-88}$ & $200^{+70}_{-69}$ & $109^{+26}_{-25}$ & $2.41^{+0.39}_{-0.49}$ & $0.29$ \\
\hline
\multirow{2}{*}{RUBIES~EGS~55604} & H$\alpha$ & $15.86^{+1.03}_{-0.96}$ & $260^{+32}_{-24}$ & $272^{+21}_{-42}$ & $69^{+21}_{-13}$ & $1.71^{+0.09}_{-0.09}$ & $0.44$ \\
 & H$\beta$ & $16.25^{+0.73}_{-0.77}$ & $158^{+103}_{-90}$ & $204^{+67}_{-70}$ & $104^{+28}_{-26}$ & $2.50^{+0.33}_{-0.47}$ & $0.22$ \\
\hline
\multirow{2}{*}{JADES~GN~68797} & H$\alpha$ & $14.30^{+0.66}_{-0.30}$ & $320^{+16}_{-15}$ & $221^{+27}_{-31}$ & $81^{+28}_{-21}$ & $1.80^{+0.05}_{-0.04}$ & $1.33$ \\
 & H$\beta$ & $14.92^{+0.37}_{-0.62}$ & $279^{+84}_{-112}$ & $191^{+72}_{-64}$ & $113^{+27}_{-37}$ & $2.54^{+0.32}_{-0.49}$ & $0.29$ \\
\hline
\multirow{2}{*}{RUBIES~EGS~42046} & H$\alpha$ & $14.71^{+0.30}_{-0.15}$ & $331^{+24}_{-21}$ & $144^{+41}_{-29}$ & $141^{+7}_{-22}$ & $1.74^{+0.08}_{-0.08}$ & $0.82$ \\
 & H$\beta$ & $17.60^{+0.35}_{-2.29}$ & $303^{+69}_{-102}$ & $193^{+71}_{-65}$ & $118^{+22}_{-30}$ & $1.65^{+0.67}_{-0.57}$ & $0.19$ \\
\hline
\multirow{2}{*}{RUBIES~UDS~154183} & H$\alpha$ & $13.92^{+0.22}_{-0.19}$ & $-57^{+20}_{-24}$ & $280^{+14}_{-25}$ & $131^{+13}_{-20}$ & $1.11^{+0.09}_{-0.08}$ & $0.23$ \\
 & H$\beta$ & $13.98^{+0.25}_{-0.22}$ & $-165^{+251}_{-131}$ & $206^{+67}_{-73}$ & $101^{+34}_{-34}$ & $2.43^{+0.40}_{-0.59}$ & $0.21$ \\
\hline
\multirow{2}{*}{GN~15498} & H$\alpha$ & $14.96^{+2.38}_{-0.95}$ & $21^{+52}_{-14}$ & $273^{+20}_{-53}$ & $95^{+40}_{-37}$ & $1.00^{+0.12}_{-0.16}$ & $0.30$ \\
 & H$\beta$ & $14.97^{+2.10}_{-0.89}$ & $-107^{+217}_{-150}$ & $198^{+70}_{-68}$ & $98^{+35}_{-33}$ & $2.38^{+0.44}_{-0.70}$ & $0.30$ \\
\hline
\multirow{2}{*}{GS~13971} & H$\alpha$ & $14.75^{+0.50}_{-0.41}$ & $119^{+27}_{-27}$ & $275^{+18}_{-37}$ & $59^{+13}_{-7}$ & $1.20^{+0.09}_{-0.09}$ & $0.32$ \\
 & H$\beta$ & $14.76^{+0.50}_{-0.42}$ & $44^{+243}_{-294}$ & $201^{+67}_{-68}$ & $100^{+35}_{-33}$ & $1.72^{+0.88}_{-0.82}$ & $0.09$ \\
\hline
\multirow{2}{*}{\textit{GS~3073}} & \textit{H$\alpha$} & $13.78^{+0.29}_{-0.98}$ & $100^{+154}_{-25}$ & $292^{+6}_{-12}$ & $139^{+8}_{-58}$ & $1.77^{+0.07}_{-0.11}$ & $0.50$ \\
 & \textit{H$\beta$} & $13.87^{+0.23}_{-0.99}$ & $-54^{+218}_{-126}$ & $254^{+33}_{-55}$ & $104^{+32}_{-35}$ & $1.34^{+0.22}_{-0.21}$ & $0.18$ \\
\hline
\multirow{2}{*}{\textit{JADES~GN~38147}} & \textit{H$\alpha$} & $14.97^{+1.41}_{-0.96}$ & $232^{+95}_{-83}$ & $155^{+80}_{-41}$ & $80^{+36}_{-21}$ & $0.78^{+0.18}_{-0.16}$ & $0.27$ \\
 & \textit{H$\beta$} & $14.94^{+1.40}_{-0.94}$ & $1^{+269}_{-267}$ & $203^{+67}_{-70}$ & $99^{+35}_{-33}$ & $1.66^{+0.91}_{-0.81}$ & $0.08$ \\
\hline
\end{tabular}}
\end{table*}

\begin{table*}
\caption{\textbf{Clumpy model fitted parameters.} As Extended Data
Table~\ref{tab:ext_smooth}, for the clumpy radiative-transfer model
(adds the covering factor $f_c$, shown as $\log\,f_c$). All other
columns and sources as in Extended Data Table~\ref{tab:ext_smooth}.
}
\label{tab:ext_clumpy}
\resizebox{\linewidth}{!}{%
\begin{tabular}{llccccccc}
\hline
Object & Fit
  & $\log\,N_{\rm HI,\mathit{n}=2}$ (cm$^{-2}$)
  & $\log\,f_c$
  & $v_{\rm exp}$ (km\,s$^{-1}$)
  & $\sigma_{\rm emit}$ (km\,s$^{-1}$)
  & $\sigma_{\rm ran}$ (km\,s$^{-1}$)
  & $\tau_{\rm Th}$
  & $\chi^2_{\rm r}$ \\
\hline
\multirow{2}{*}{GN~9771} & H$\alpha$ & $15.71^{+0.27}_{-0.43}$ & $0.83^{+0.05}_{-0.05}$ & $169^{+6}_{-6}$ & $244^{+10}_{-9}$ & $91^{+3}_{-3}$ & $1.80^{+0.02}_{-0.02}$ & $1.15$ \\
 & H$\beta$ & $15.75^{+0.24}_{-0.35}$ & $1.49^{+0.30}_{-0.22}$ & $178^{+48}_{-44}$ & $229^{+48}_{-86}$ & $97^{+21}_{-21}$ & $2.88^{+0.08}_{-0.14}$ & $0.39$ \\
 \hline
 \multirow{2}{*}{A2744~45924} & H$\alpha$ & $15.27^{+0.76}_{-1.37}$ & $1.26^{+0.45}_{-0.09}$ & $-28^{+5}_{-12}$ & $111^{+83}_{-10}$ & $144^{+5}_{-17}$ & $1.89^{+0.01}_{-0.01}$ & $13.88$ \\
 & H$\beta$ & $16.08^{+0.15}_{-0.12}$ & $1.87^{+0.10}_{-0.14}$ & $-106^{+23}_{-23}$ & $257^{+28}_{-47}$ & $92^{+8}_{-8}$ & $2.90^{+0.07}_{-0.12}$ & $2.13$ \\
\hline
\multirow{2}{*}{RUBIES~EGS~49140} & H$\alpha$ & $18.24^{+0.40}_{-0.82}$ & $1.50^{+0.31}_{-0.30}$ & $16^{+10}_{-9}$ & $217^{+59}_{-80}$ & $127^{+17}_{-30}$ & $1.79^{+0.13}_{-0.15}$ & $0.30$ \\
 & H$\beta$ & $18.33^{+0.38}_{-0.60}$ & $1.56^{+0.30}_{-0.36}$ & $-84^{+89}_{-89}$ & $199^{+72}_{-70}$ & $117^{+22}_{-28}$ & $2.49^{+0.34}_{-0.46}$ & $0.29$ \\
\hline
\multirow{2}{*}{RUBIES~EGS~55604} & H$\alpha$ & $16.60^{+0.83}_{-1.25}$ & $1.75^{+0.18}_{-0.28}$ & $256^{+33}_{-30}$ & $272^{+21}_{-70}$ & $69^{+12}_{-11}$ & $1.77^{+0.10}_{-0.10}$ & $0.50$ \\
 & H$\beta$ & $16.91^{+0.63}_{-0.95}$ & $1.57^{+0.29}_{-0.33}$ & $170^{+106}_{-92}$ & $203^{+69}_{-70}$ & $107^{+27}_{-27}$ & $2.53^{+0.31}_{-0.45}$ & $0.23$ \\
\hline
\multirow{2}{*}{JADES~GN~68797} & H$\alpha$ & $15.17^{+2.47}_{-0.86}$ & $1.47^{+0.45}_{-0.87}$ & $314^{+22}_{-33}$ & $168^{+46}_{-51}$ & $62^{+19}_{-9}$ & $1.84^{+0.07}_{-0.06}$ & $1.88$ \\
 & H$\beta$ & $17.44^{+0.33}_{-2.49}$ & $1.40^{+0.39}_{-0.52}$ & $288^{+79}_{-117}$ & $197^{+70}_{-66}$ & $108^{+29}_{-34}$ & $2.45^{+0.38}_{-0.50}$ & $0.28$ \\
\hline
\multirow{2}{*}{RUBIES~EGS~42046} & H$\alpha$ & $16.16^{+1.08}_{-1.19}$ & $1.46^{+0.23}_{-0.19}$ & $347^{+28}_{-30}$ & $116^{+25}_{-12}$ & $135^{+10}_{-22}$ & $1.76^{+0.09}_{-0.08}$ & $0.73$ \\
 & H$\beta$ & $16.89^{+0.64}_{-0.88}$ & $1.71^{+0.21}_{-0.32}$ & $313^{+63}_{-104}$ & $201^{+66}_{-68}$ & $123^{+19}_{-30}$ & $1.78^{+0.64}_{-0.54}$ & $0.20$ \\
\hline
\multirow{2}{*}{RUBIES~UDS~154183} & H$\alpha$ & $15.51^{+1.97}_{-1.44}$ & $1.20^{+0.46}_{-0.40}$ & $-53^{+17}_{-21}$ & $272^{+20}_{-31}$ & $105^{+23}_{-25}$ & $1.05^{+0.10}_{-0.12}$ & $0.24$ \\
 & H$\beta$ & $16.86^{+1.02}_{-1.66}$ & $1.41^{+0.40}_{-0.57}$ & $-170^{+105}_{-110}$ & $206^{+67}_{-73}$ & $107^{+30}_{-34}$ & $2.05^{+0.62}_{-0.69}$ & $0.15$ \\
\hline
\multirow{2}{*}{GN~15498} & H$\alpha$ & $17.62^{+0.88}_{-3.13}$ & $1.57^{+0.30}_{-0.37}$ & $31^{+34}_{-16}$ & $254^{+34}_{-81}$ & $95^{+29}_{-29}$ & $0.86^{+0.17}_{-0.19}$ & $0.26$ \\
 & H$\beta$ & $17.75^{+0.80}_{-2.71}$ & $1.03^{+0.62}_{-0.63}$ & $-100^{+250}_{-161}$ & $203^{+67}_{-71}$ & $100^{+35}_{-34}$ & $2.40^{+0.42}_{-0.70}$ & $0.30$ \\
\hline
\multirow{2}{*}{GS~13971} & H$\alpha$ & $15.47^{+1.33}_{-0.86}$ & $1.51^{+0.32}_{-0.31}$ & $122^{+24}_{-25}$ & $275^{+18}_{-34}$ & $54^{+6}_{-3}$ & $1.20^{+0.10}_{-0.10}$ & $0.33$ \\
 & H$\beta$ & $15.41^{+1.37}_{-0.84}$ & $1.02^{+0.66}_{-0.67}$ & $29^{+252}_{-291}$ & $200^{+68}_{-68}$ & $99^{+34}_{-33}$ & $1.73^{+0.86}_{-0.82}$ & $0.09$ \\
\hline
\multirow{2}{*}{\textit{GS~3073}} & \textit{H$\alpha$} & $14.23^{+1.94}_{-0.52}$ & $1.33^{+0.50}_{-0.51}$ & $91^{+27}_{-22}$ & $287^{+10}_{-20}$ & $121^{+16}_{-16}$ & $1.78^{+0.06}_{-0.07}$ & $0.47$ \\
 & \textit{H$\beta$} & $14.28^{+1.65}_{-0.53}$ & $1.01^{+0.66}_{-0.65}$ & $-31^{+141}_{-131}$ & $246^{+38}_{-63}$ & $102^{+33}_{-35}$ & $1.30^{+0.23}_{-0.23}$ & $0.18$ \\
\hline
\multirow{2}{*}{\textit{JADES~GN~38147}} & \textit{H$\alpha$} & $15.72^{+1.65}_{-1.75}$ & $1.21^{+0.51}_{-0.64}$ & $251^{+89}_{-120}$ & $139^{+63}_{-29}$ & $86^{+36}_{-25}$ & $0.79^{+0.19}_{-0.17}$ & $0.30$ \\
 & \textit{H$\beta$} & $15.67^{+1.69}_{-1.77}$ & $0.99^{+0.68}_{-0.65}$ & $-2^{+269}_{-272}$ & $205^{+66}_{-71}$ & $99^{+35}_{-34}$ & $1.69^{+0.89}_{-0.84}$ & $0.08$ \\
\hline
\\
\end{tabular}}
\end{table*}

\end{appendices}

\end{document}